\documentclass[11pt]{article}

\catcode`\@=11
\@addtoreset{equation}{section}

\evensidemargin \oddsidemargin

\usepackage{mathrsfs,amsbsy,amssymb,latexsym,amsfonts,amsmath,
amssymb,
fancyhdr,lastpage,
fancybox,
ulem,cases
,wrapfig,}

\usepackage{cite}

  \usepackage{ascmac}
  
  \usepackage{bm}

\newcommand\CW{\mathcal{W}}
\newcommand\CS{\mathcal{S}}
\newcommand\CR{\mathcal{R}}

\newcommand\msD{\mathscr{D}}
\newcommand\msS{\mathscr{S}}
\newcommand\msR{\mathscr{R}}
\newcommand\msG{\mathscr{G}}
\newcommand\msF{\mathscr{F}}

\newcommand\msSsl{{\msS\hspace{-2.5mm}\slash\,}}
\newcommand\msRsl{{\msR\hspace{-2.5mm}\slash\,}}
\newcommand\msRt{{\tilde\msR}}
\newcommand\msGt{{\tilde\msG}}

\newcommand\SDsl{\mathscr{D}\hspace{-2.1mm}/}
\newcommand\Asl{A\hspace{-2.1mm}/}

\newcommand\pa{\partial}

\newcommand\gb{\boldsymbol{g}}

\newcommand\CSsl{{\CS\hspace{-2.4mm}\slash\,}}
\newcommand\zetasl{{\zeta\hspace{-2.0mm}\slash\,}}
\newcommand\CRsl{{\CR\hspace{-2.7mm}\slash\,}}

\newcommand\intd{\int\!\d^Dx\,}

\newcommand\nn{\nonumber}

\newcommand\disp{\displaystyle}

\newcommand\phis{{\phi^*}}
\newcommand\psis{{\psi^*}}

\newcommand\tr{{\rm tr}\,}

\renewcommand\d{{\rm d}}
\newcommand\diag{{\rm diag}}

\newcommand\bref[1]{(\ref{#1})}

\newcommand\pat{\tilde\pa}
\newcommand\Gt{\tilde G}
\newcommand\CRt{\tilde \CR}
\newcommand\pasl{{\partial\hspace{-2.2mm}\slash\,}}
\newcommand\psisl{{\psi\hspace{-1.8mm}\slash}}
\newcommand\epsilonsl{{\epsilon\hspace{-1.5mm}\slash}}

\begin{document}

\hfill 
\vspace{3cm}

\begin{center}
{\LARGE \bf
Higher-Spin  
Gauge Models
\\\vspace{5mm}
in the BRST-antifield Formalism
}
\vspace{30mm}\\
Ryota Fujii,~
Hiraki Kanehisa,~
Makoto Sakaguchi
\,and~
Haruya Suzuki

\vspace{15mm}

Department of Physics, Ibaraki University, Mito 310-8512, Japan
\end{center}

\vspace{25mm}

\begin{abstract}

We examine the mass shell condition of the bosonic higher spin
gauge models which are BRST deformations of the Fronsdal action
in the BRST-antifield formalism.
Assuming convergence of the infinite series of the deformation parameter, 
we will find that these models are free on-shell.
We further investigate whether this nature persists for models with fermions and models on AdS spaces.
We find that these models also turn out to be free on-shell after all.
Furthermore, we point out that the cubic vertices of these models are BRST-exact.

\end{abstract}

\thispagestyle{empty}
\setcounter{page}{0}

\newpage

\tableofcontents

\setcounter{footnote}{0}

\section{Introduction}
Higher-spin gauge theories are expected to reveal characteristic features of string theory in the high-energy limit \cite{Gross} (see also \cite{Sagnotti tensionless limit}).
They also provided nontrivial examples of the AdS/CFT correspondence \cite{AdS/CFT}.
A bosonic higher-spin gauge theory on a three-dimensional AdS space (AdS$_3$)\,\cite{3dVasiliev}
has been conjectured \cite{3dV/W} to be dual to  
the 't Hooft limit of the $\CW_N$ minimal model,
while Vasiliev's higher-spin gauge theory on 
AdS$_4$\,\cite{4dVasiliev}
is conjectured\,\cite{4dV/O} to be dual to the three-dimensional $O(N)$ sigma-model.

Free higher-spin gauge theories are now well understood.
Interacting theories of massless higher-spin gauge fields in flat spacetime
are severely constrained by the no-go theorem \cite{no-go}\footnote{See \cite{KU81} for an analysis in the field theory framework.}. 
To avoid this, spin values and the number of derivatives contained in interaction vertices
should be restricted so that the higher-spin gauge fields are not included in the asymptotic states. 
By using the light-cone formulation\footnote{
Remarkably,
chiral higher-spin gravity \cite{Met91,PS1609} in four-dimensions is shown to 
be one-loop finite \cite{STT2002,ST2004}.
See also \cite{P1710,KST2105}.
}, 
a classification of spin values and the number of derivatives of allowed cubic vertices
for totally-symmetric bosonic and fermionic fields
 was given in \cite{Metsaev0512,Metsaev0712}.
In constructing interaction vertices explicitly\footnote{
See \cite{Bonora2020} for a different approach to generalize YM theory or gravity.},
the Noether's procedure is used in
\cite{BBvD85,MMR Noether,Sleight1704,KMP2104},
and the BRST approach to impose gauge  invariances
is used in
\cite{PT9803,BPT0101,BKP0410,BK0505,BKRT0603,BR2010,BKTW2103,BR2105} on flat spacetime
and  \cite{BFPT0609,FIPT0708} on AdS spaces (see  \cite{BCS0409,FT0805} for review).

In this paper, we employ the BRST-antifield formalism \cite{BH9304}
which is known to be very powerful in constructing interactions systematically.
Using this cohomological method\footnote{See \cite{ BL2104} for a different approach.}, 
bosonic interaction vertices are constructed in
\cite{BDGH0007, BRST cohomology}
and those including gauge fermions in
\cite{HLGR1206,HLGR1310, Rahman1905}.
The results in \cite{HLGR1206,HLGR1310} are shown to be consistent with the classification \cite{Metsaev0512,Metsaev0712}
and with those obtained in the tensionless limit of open string theory \cite{Sagnotti tensionless limit}
(see also \cite{ST0311,FT0705,FT1009,Polyakov0910,Sagnotti1112}).

\bigskip

In \cite{SS2011},
by applying the BRST deformation scheme,
 bosonic higher-spin gauge models of massless totally-symmetric tensors
 are constructed.
These models contain vertices of all orders in the deformation parameter.
It is noted that the action $S$ satisfies the master equation $(S,S)=0$ and so is BRST-invariant exactly.
In constructing vertices,
the Fronsdal tensor \cite{Fronsdal78} is employed as a building block.
Since the Fronsdal tensor may be defined on anti-de Sitter (AdS) space,
these models constructed on flat space are generalized to those on AdS spaces naturally.
In the present paper,
we see that the BRST deformation scheme
can be formulated
on AdS space also.

One of the purposes of this paper is to examine the mass shell condition of the bosonic higher spin
gauge models which are gneralizations of the model given in  \cite{SS2011}.
Assuming convergence of the infinite series of the deformation parameter, 
we will find that these models are free on-shell.
We further investigate whether this nature persists for models with fermions and models on AdS spaces.
We find that these models also turn out to be free on-shell after all.
Furthermore, we point out that the cubic vertices of these models are BRST-exact.
Further work is needed to determine whether these models are also free off-shell.

Introducing massless bosonic totally-symmetric tensors,
we construct higher spin gauge actions which are generalizations of the model given in \cite{SS2011}
by applying BRST deformation sheme in section 3.
Further
we introduce massless totally-symmetric rank-$n$ tensor-spinors,
which are Dirac spinors in $D$-dimensional spacetime.
By applying the BRST deformation scheme,
a free higher-spin gauge action $S$ is deformed such that the deformed action would satisfy the master equation.
In section 4,
we examine the models with two fermions and one boson,
while,
in section 5,
the models with a fermion and a boson are examined.
As a result, we obtain the actions including all order vertices for each case.
These actions satisfy the master equation exactly.
Furthermore,
while each vertex considered in \cite{SS2011} forms an open chain of fields,
in this paper,
more general case,
in which each vertex forms a closed chain of fields, is considered.

\medskip

The paper is organized as follows.
In the next section, we introduce free higher-spin gauge theories of a bosonic tensor 
and  a fermionic tensor-spinor in the BRST-antifield formalism.
The BRST deformation scheme is explained as well.
After deriving higher-spin gauge models of three bosons in section \ref{sec:three bosonic gauge firleds},
higher-spin gauge model of two fermions and one boson
is derived in section \ref{sec:three fields},
and that of a fermion and a boson is derived in section \ref{sec:two fields}.
After introducing objects on AdS spaces,
we derive higher-spin gauge models on AdS spaces in the BRST deformation scheme in section \ref{sec:AdS}.
In each case, we examine the mass shell conditions to see if they are free on-shell.
The last section is devoted to a summary and discussions.
In appendix A, 
we present higher-spin gauge models
of two bosons on flat and AdS spaces which are generalization of the models found in \cite{SS2011}.

\section{Free higher-spin gauge theory and BRST deformation scheme }
\label{sec:free}

We explain the notations used throughout this paper.
We will introduce gauge fermions as well as gauge bosons
in $D$-dimensional flat spacetime with $D\geq 4$.
Free actions
of higher-spin gauge fields are formulated  in the BRST-antifield formalism.
We explain the BRST deformation scheme used in constructing vertices.

\subsection{Bosonic higher-spin gauge theory}
\label{sec:free bos}
First we introduce a totally-symmetric rank-$n$ tensor bosonic gauge field  $\phi_{\mu_1\cdots\mu_n}$
in  $D$-dimensional flat spacetime.
The Fronsdal tensor \cite{Fronsdal78} for this field
is defined by\footnote{
Our convention for symmetrization of indices is
\begin{align}
\pa_{(\mu_1}\cdots\pa_{\mu_r}\phi_{\mu_{r+1}\cdots\mu_n)\nu_1\cdots\nu_m}
=\frac{1}{r! (n-r)!}\sum_{\{\mu_1,\ldots,\mu_n\}}
\pa_{\mu_1}\cdots\pa_{\mu_r}\phi_{\mu_{r+1}\cdots\mu_n\nu_1\cdots\nu_m}
\nn,
\end{align}
where $\{\mu_1,\ldots,\mu_n\}$ indicates that 
the sum is taken over permutations of $\mu_1,\ldots,\mu_n$.
}
\begin{align}
F_{\mu_1\cdots\mu_n}(\phi)=&\square\phi_{\mu_1\cdots\mu_n}
-\pa_{(\mu_1}\pa\cdot\phi_{\mu_2\cdots\mu_n)}
+\pa_{(\mu_1}\pa_{\mu_2}\phi'_{\mu_3\cdots\mu_n)},
\label{Fronsdal bos flat}
\end{align}
where $\square=\eta^{\mu\nu}\pa_\mu\pa_\nu$
and $\eta^{\mu\nu}=\diag(-1,+1,\ldots,+1)$.
A prime on a field represents the trace, 
namely $\phi'_{\mu_3\cdots \mu_n}=\phi_{\mu_3\cdots \mu_n}{}^\rho{}_\rho$\,.
The divergence of $\phi$
is expressed as
 $\pa\cdot\phi_{\mu_2\cdots\mu_n}=\pa^{\mu_1}\phi_{\mu_1\mu_2\cdots\mu_n}$\,.
Under the gauge transformation
with a totally-symmetric rank-$(n-1)$ tensor parameter $\xi_{\mu_2\cdots \mu_n}$,
\begin{align}
\delta \phi_{\mu_1\cdots\mu_n}
=\pa_{(\mu_1}\xi_{\mu_2\cdots\mu_n)},
\label{eqn:gauge trans bosnic}
\end{align}
the Fronsdal tensor $F_{\mu_1\cdots\mu_n}$
is transformed to
$
\delta F_{\mu_1\cdots\mu_n}=3\pa_{(\mu_1}\pa_{\mu_2}\pa_{\mu_3}\xi'_{\mu_4\cdots\mu_n)}
$.
Here and hereafter,
we require that the gauge parameter $\xi$ is traceless for $n\geq 3$
\begin{align}
\xi'_{\mu_4\cdots\mu_n}=0,
\end{align}
which ensures the gauge invariance of 
the Fronsdal tensor.
Furthermore we
impose
a double traceless constraint on $\phi$ for $n\geq 4$
\begin{align}
\phi''_{\mu_5\cdots\mu_n}=0,
\label{eqn:phi''=0}
\end{align}
so that the Fronsdal equation $F(\phi)=0$\footnote{We frequently omit totally-symmetric indices
but this may not cause confusion.}
should describe
the propagation of a massless rank-$n$ tensor gauge field.
The action which leads to the Fronsdal equation
is
\begin{align}
S_\phi&=
\intd \frac{1}{2}\phi^{\mu_1\cdots\mu_n}G_{\mu_1\cdots\mu_n},
\label{eqn:action phi}
\end{align}
where $G_{\mu_1\cdots\mu_n}$ is defined by
\begin{align}
G_{\mu_1\cdots\mu_n}(\phi)&\equiv
F_{\mu_1\cdots\mu_n}-\frac{1}{2}\eta_{(\mu_1\mu_2}F'_{\mu_3\cdots\mu_n)}.
\label{eqn:G}
\end{align}
Varying the action $S_\phi$ with respect to $\phi$, we obtain 
the equation of motion $G_{\mu_1\cdots\mu_n}=0$.
A useful relation to derive this equation is\footnote{We drop surface terms throughout this paper.
}
\begin{align}
\intd \varphi_1^{\mu_1\cdots\mu_n}\tilde G_{\mu_1\cdots\mu_n}(\varphi_2)
&=\intd \tilde G_{\mu_1\cdots\mu_n}(\varphi_1)\varphi_2^{\mu_1\cdots\mu_n},
\label{eqn:relation Gt}
\end{align}
where $\varphi_i$ are arbitrary totally-symmetric rank-$n$
tensors.
We have defined $\tilde G(A)$
by
\begin{align}
\tilde G_{\mu_1\cdots\mu_n}(A)\equiv &
G_{\mu_1\cdots\mu_n}(A)+\frac{1}{2}\eta_{(\mu_1\mu_2}\pa_{\mu_3}\pa_{\mu_4}A''_{\mu_5\cdots\mu_n)},
\label{eqn:Gt}
\end{align}
where $A$ is an arbitrary totally-symmetric rank-$n$ tensor.
When the  double traceless condition for $\varphi_i$
would be implemented, 
one obtains 
$\tilde G(\varphi_i)=G(\varphi_i)$.
Using the relation \bref{eqn:relation Gt} with $\Gt=G$, 
we obtain  $G(\phi)=0$ from the variation of \bref{eqn:action phi}\footnote{
Taking the trace of $G=0$ leads to
$-\frac{1}{2}(D+2n-6)F'-\frac{1}{2}\eta F''=0$.
Assume that $D+2n-6\neq 0$,
and then we obtain $F'=0$ as $F''=0$ follows from $\phi''=0$.
This implies that $F=0$.
$D+2n-6= 0$ is satisfied for $(D,n)=(6,0)$ and $(4,1)$ as $D\geq 4$.
In both cases, $G=0$ means $F=0$ as $F'=0$.
As a result, the equation of motion $G=0$ implies the Fronsdal equation $F=0$.
}.
Similarly the gauge variation of $S_\phi$ vanishes due to the gauge invariance
$G(\delta \phi)=0$.

\medskip

In the BRST-antifield formalism,
corresponding to the gauge parameter $\xi_{\mu_2\cdots\mu_{n}}$,
we introduce a Grassmann-odd ghost field $c_{\mu_2\cdots\mu_{n}}$
with the same algebraic symmetry.
The $c_{\mu_2\cdots\mu_{n}}$ must be traceless 
$c'_{\mu_4\cdots\mu_{n}}=0$
because of the traceless condition $\xi'=0$.
The gauge invariance of $S_\phi$ is encoded to
the BRST invariance under the BRST transformation
\begin{align}
\delta_B \phi_{\mu_1\cdots\mu_n}&=\pa_{(\mu_1}c_{\mu_2\cdots\mu_n)},~~~
\delta_B c_{\mu_2\cdots\mu_n}=0.
\label{eqn:BRST phi c}
\end{align}
The gauge field $\phi$ and the ghost field $c$ are collectively called ``fields'' and denoted as
 $\Phi^A
$. 
\begin{table}[ht]
\begin{align*}
\begin{array}{cccc}\hline 
Z & pgh(Z) & agh(Z) & gh(Z) \\\hline 
\phi_{\mu_1\cdots\mu_n} & 0 & 0 & 0 \\
c_{\mu_2\cdots\mu_{n}} & 1 & 0 & 1 \\
\phi^*_{\mu_1\cdots\mu_n} & 0 & 1 & -1 \\
c^*_{\mu_2\cdots\mu_{n}} & 0 & 2 & -2 \\\hline 
\end{array}
\end{align*}
 \caption{Grading properties of fields and antifields}
 \label{table:grading bos}
\end{table}
We further introduce antifields $\Phi^*
_A=\{\phi^*_{\mu_1\cdots\mu_n},
c^*_{\mu_2\cdots\mu_{n}}\}$
which have the same algebraic symmetries but opposite Grassmann parity.
Two gradings are introduced.
One is the pure ghost number $pgh$, and the other is the antighost number $agh$.
The ghost number $gh$ is defined as $gh\equiv pgh-agh$\,.
The grading property is summarized
in Table \ref{table:grading bos}.

The antibracket
for two functionals, $X(\Phi^A,\Phi^*_A)$ and $Y(\Phi^A,\Phi^*_A)$,
is defined by
\begin{align}
(X,Y)\equiv 
X\frac{\stackrel{\leftarrow}{\delta}}{\delta\Phi^A}\frac{\stackrel{\rightarrow}{\delta}}{\delta\Phi^*_A}Y
-X\frac{\stackrel{\leftarrow}{\delta}}{\delta\Phi^*_A}\frac{\stackrel{\rightarrow}{\delta}}{\delta\Phi^A}Y.
\end{align}
The action $S_\phi$
in \bref{eqn:action phi}
can be extended to $S^0[\Phi,\Phi^*]$ such that
the BRST transformation of a functional $X(\Phi^A,\Phi^*_A)$
is expressed as
\begin{align}
\delta_B X=(X,S^0)
\,.
\end{align}
Note that $\delta_B$ acts from the right.
The nilpotency $\delta_B^2=0$
requires the master equation $(S^0,S^0)
=0$\,.
In the present case,  $S^0$ may be given as
\begin{align}
S^0[\Phi,\Phi^*]=S_\phi+\int \d ^Dx\, 
\phis{}^{\mu_1\cdots \mu_n}\pa_{(\mu_1}c_{\mu_2\cdots \mu_n)}
,
\label{eqn:S^0 bos}
\end{align}
which leads to
\bref{eqn:BRST phi c}
and
\begin{align}
\delta_B\phi^*_{\mu_1\cdots\mu_n}&=-G_{\mu_1\cdots\mu_n},~~~
\label{eqn:BRST phi*}\\
\delta_B c^*_{\mu_2\cdots\mu_n}&=
-n\pa\cdot\phi^*_{\mu_2\cdots\mu_n}
+
\frac{n}{D+2n-6}
\eta_{(\mu_2\mu_3}\pa\cdot\phis{}'_{\mu_4\cdots\mu_n)}
,
\label{eqn:BRST c*}
\end{align}
when $D+2n-6\neq 0$\footnote{
We comment on the case $D+2n-6=0$,
namely $(D,n)=(6,0)$ and $(4,1)$.
When $(D,n)=(6,0)$ the gauge transformation \bref{eqn:gauge trans bosnic} becomes trivial,
and so ghost and anti-fields are not introduced.
When $(D,n)=(4,1)$, the second term on the right-hand side of \bref{eqn:BRST c*}
is not needed
because $\pa^\mu G_\mu=\pa^\mu(\square \phi_\mu-\pa_\mu\pa\cdot\phi)=0$. 
}.
The last term on the right-hand side of \bref{eqn:BRST c*}
is required for the nilpotency  $\delta_B^2 c^*=0$.
We note that the key relation for the nilpotency,
$
-\pa\cdot G_{\mu_2\cdots\mu_n}
+
\frac{1}{D+2n-6}
\eta_{(\mu_2\mu_3}\pa\cdot G'_{\mu_4\cdots\mu_n)}
=0
$,
follows from the double traceless condition \bref{eqn:phi''=0}.
Since $\delta_B c^*=(c^*,S^0)$,
the last term on the right-hand side of \bref{eqn:BRST c*}
requires an additional term,
$
\int \d^Dx\,\frac{n}{D+2n-6}
\eta_{(\mu_2\mu_3}\pa\cdot\phis{}'_{\mu_4\cdots\mu_n)}c^{\mu_2\cdots\mu_n}
$,
in the action $S^0$.
However this term disappears
 due to $c'=0$,
and
leaves
the action \bref{eqn:S^0 bos}
unchanged.

\subsection{Fermionic higher-spin  gauge theory}  
\label{sec:free fer}

Next, we introduce a totally-symmetric rank-$n$ tensor-spinor gauge field $\psi_{\mu_1\cdots\mu_n}$
which
is a Dirac spinor
in  $D$-dimensional flat spacetime.
The Fronsdal tensor \cite{FangFronsdal78} for this field 
is defined by
\begin{align}
\CS_{\mu_1\cdots\mu_n}(\psi)
=&
i\left(\pasl\psi_{\mu_1\cdots\mu_n}
-\pa_{(\mu_1}\psisl_{\mu_2\cdots\mu_n)}
\right).
\label{eqn:Fronsdal fer flat}
\end{align}
A slash of an object denotes the $\gamma$-trace of the object,
namely
$\pasl=\gamma^\mu\pa_\mu$ and 
$\psisl_{\mu_2\cdots\mu_n}=\gamma^{\mu_1}\psi_{\mu_1\cdots\mu_n}$.
The Dirac matrices $\gamma^\mu$
satisfy
$
\{\gamma^\mu,\gamma^\nu\}=2\eta^{\mu\nu}
$.
We choose $\gamma^0$ as anti-hermite $\gamma^0{}^\dag=-\gamma^0$
while $\gamma^i$ as hermite $\gamma^i{}^\dag=\gamma^i$,
so that
$
\gamma^\mu{}^\dag \gamma^0=-\gamma^0 \gamma^\mu
$.
The Dirac-conjugate of  a spinor $\psi$ is defined by $\bar\psi \equiv \psi^\dag \gamma^0$.
Our convention  of the hermite conjugate is that
$(\bar \varphi_1 \varphi_2)^\dag=-\bar\varphi_2 \varphi_1$,
where $\varphi_i$ are either Grassmann even spinor or Grassmann odd spinor.

Under the gauge transformation
with a totally-symmetric rank-$(n-1)$ tensor-spinor parameter
$\epsilon_{\mu_2\cdots\mu_n}$,
\begin{align}
\delta\psi_{\mu_1\cdots\mu_n}
=\pa_{(\mu_1}\epsilon_{\mu_2\cdots\mu_n)},
\label{eqn:gauge trans fermion}
\end{align}
the Fronsdal tensor 
$\CS$ is transformed to
$
\delta \CS_{\mu_1\cdots\mu_n}=-2i\pa_{(\mu_1}\pa_{\mu_2} \epsilonsl_{\mu_3\cdots\mu_n)}
$.
Here and hereafter,
we require that $\epsilon$ is $\gamma$-traceless for $n\geq 2$
\begin{align}
\epsilonsl_{\mu_3\cdots\mu_n}=0,
\end{align}
which ensures the gauge invariance of 
the Fronsdal tensor $\CS$.
Furthermore we
impose
a triple $\gamma$-traceless constraint on $\psi$ for  $n\geq 3$
\begin{align}
\psisl'_{\mu_4\cdots\mu_n}=0,
\label{psisl'=0}
\end{align}
so that the Fronsdal equation
$\CS(\psi)=0$ should describe
the propagation of a massless rank-$n$ tensor-spinor gauge  field.
The action which leads to the Fronsdal equation
 $\CS=0$ 
is  given by
\begin{align}
S_\psi
&=\int \d ^Dx\, \frac{1}{2}\left[
\bar \CR_{\mu_1\cdots\mu_n}\psi^{\mu_1\cdots\mu_n}
-
\bar\psi^{\mu_1\cdots\mu_n}\CR_{\mu_1\cdots\mu_n}
\right],
\label{eqn:action psi}
\end{align}
where we introduced 
$\CR_{\mu_1\cdots\mu_n}$ as
\begin{align}
\CR_{\mu_1\cdots\mu_n}(\psi)&\equiv
\CS_{\mu_1\cdots\mu_n}
-\frac{1}{2}\gamma_{(\mu_1}\CSsl_{\mu_2\cdots\mu_n)}
-\frac{1}{2}\eta_{(\mu_1\mu_2} \CS'_{\mu_3\cdots\mu_n)}.
\label{eqn:R}
\end{align}
A useful and impotant relation we frequently use throughout this paper
is
\begin{align}
\int\d^D x\, \bar{\tilde\CR}_{\mu_1\cdots\mu_n}(\psi_1)\psi_2^{\mu_1\cdots\mu_n}
&=-\int\d^D x\, \bar\psi_1^{\mu_1\cdots\mu_n} \tilde\CR_{\mu_1\cdots\mu_n}(\psi_2)
.
\label{eqn:relation Rt}
\end{align}
where
$\psi_i$ are arbitrary totally-symmetric rank-$n$ tensor-spinors.
We have defined  
$\CRt(A)$ by
\begin{align}
\CRt_{\mu_1\cdots\mu_n}(A)\equiv& 
\CR_{\mu_1\cdots\mu_n}(A)
-\frac{i}{2}\eta_{(\mu_1\mu_2}\pa_{\mu_3}A\hspace{-2.2mm}/ '_{\mu_4\cdots\mu_n)},
\end{align}
where $A$ is an arbitrary totally-symmetric rank-$n$ tensor-spinor.
When
the triple $\gamma$-traceless condition for $\psi_i$ 
would be implemented, 
one has
$\CRt(\psi_i)=\CR(\psi_i)$.
Varying the action $S_\psi$ with respect to $\bar\psi$
and using \bref{eqn:relation Rt}, we obtain 
the equation of motion $\CR_{\mu_1\cdots\mu_n}=0$\,.
Taking the $\gamma$-trace of $\CR=0$, we obtain $-\frac{1}{2}(D+2n-4)\CSsl-\frac{1}{2}\eta \CSsl'=0$.
When $D+2n-4\neq0$, 
this implies that  $\CSsl=0$,
since $\CSsl'=0$ follows from $\psisl'=0$.
Further taking the $\gamma$-trace of $\CSsl=0$,
we obtain $\CS'=0$.
These imply that $\CS=0$.
When $D+2n-4=0$,
$\CR=\CS$ follows because $(D,n)=(4,0)$ for $D\geq 4$ and $n\geq0$.
This is because $\CR=0$ means $\CS=0$. 
As a result, the equation of motion $\CR=0$ implies the Fronsdal equation $\CS=0$.
Similarly the gauge invariance of $S_\psi$
is ensured by the gauge invariance $\CR(\delta\psi)=0$.

\medskip

Corresponding to the tensor-spinor gauge parameter $\epsilon_{\mu_2\cdots\mu_{n}}$,
we introduce a Grassmann-even ghost tensor-spinor $\zeta_{\mu_2\cdots\mu_{n}}$
with the same algebraic symmetry.
The $\zeta_{\mu_2\cdots\mu_{n}}$ must be $\gamma$-traceless $\zetasl_{\mu_3\cdots\mu_{n}}=0$,
just like $\epsilonsl_{\mu_3\cdots\mu_{n}}=0$.
The gauge invariance of $S_\psi$ is encoded to
the BRST invariance under the BRST transformation
\begin{align}
\delta_B \psi_{\mu_1\cdots\mu_n}&=\pa_{(\mu_1}\zeta_{\mu_2\cdots\mu_n)},~~~
\delta_B \zeta_{\mu_2\cdots\mu_n}=0.
\label{BRST psi zeta}
\end{align}
In addition to spinor fields  $\Psi^A\equiv 
\{\psi_{\mu_1\cdots\mu_n},~\zeta_{\mu_2\cdots\mu_{n}}\}$,
we introduce spinor antifields $\Psi^*
_A\equiv \{\psi^*_{\mu_1\cdots\mu_n},
\zeta^*_{\mu_2\cdots\mu_{n}}\}$
which have the same algebraic symmetries but opposite Grassmann parity.
The grading property is summarized
in Table \ref{table:grading fer}.
\begin{table}[ht]
\begin{align*}
\begin{array}{cccc}\hline 
Z & pgh(Z) & agh(Z) & gh(Z) \\\hline 
\psi_{\mu_1\cdots\mu_n} & 0 & 0 & 0 \\
\zeta_{\mu_2\cdots\mu_{n}} & 1 & 0 & 1 \\ 
\psi^*_{\mu_1\cdots\mu_n} & 0 & 1 & -1 \\ 
\zeta^*_{\mu_2\cdots\mu_{n}} & 0 & 2 & -2 \\\hline 
\end{array}
\end{align*}
 \caption{Grading properties of fiedls and antifields}
 \label{table:grading fer}
\end{table}

The action $S_\psi$
in \bref{eqn:action psi}
can be extended to $S^0[\Psi,\Psi^*]$ such that
the BRST transformation of a functional $X(\Psi^A,\Psi^*_A)$
is expressed as
$\delta_B X=(X,S^0)$.
The antibracket
for two functionals, $X(\Psi^A,\Psi^*_A)$ and $Y(\Psi^A,\Psi^*_A)$,
is defined by
\begin{align}
(X,Y)\equiv 
X\frac{\stackrel{\leftarrow}{\delta}}{\delta\Psi^A}
\frac{\stackrel{\rightarrow}{\delta}}{\delta\bar \Psi^*_A}Y
-X\frac{\stackrel{\leftarrow}{\delta}}{\delta\Psi^*_A}
\frac{\stackrel{\rightarrow}{\delta}}{\delta\bar \Psi^A}Y.
\end{align}
In the present case,  $S^0$ may be given as
\begin{align}
S^0[\Psi,\Psi^*]=
S_\psi
+\int \d ^Dx\, \left(
\bar\psis{}^{\mu_1\cdots\mu_n}\pa_{(\mu_1}\zeta_{\mu_2\cdots\mu_n)}
-\pa_{(\mu_1}\bar\zeta_{\mu_2\cdots\mu_n)}\psis{}^{\mu_1\cdots\mu_n}
\right),
\label{eqn:S^0}
\end{align}
which leads to
\bref{BRST psi zeta}
and
\begin{align}
\delta_B\psi^*_{\mu_1\cdots\mu_n}&=\CR_{\mu_1\cdots\mu_n},~~~
\label{BRST phi*}\\
\delta_B \zeta^*_{\mu_2\cdots\mu_n}&=
-n\pa\cdot\psi^*_{\mu_2\cdots\mu_n}
+
\frac{n}{D+2n-4}
(\gamma\pa\cdot\psisl^*+\eta \pa\cdot\psis{}')
,
\label{BRST zeta*}
\end{align}
when $D+2n-4\neq 0$\footnote{
When $D+2n-4=0$, namely $(D,n)=(4,0)$,
the gauge transformation \bref{eqn:gauge trans fermion} becomes trivial,
and so a ghost and antifields are not introduced.
}.
Note that the second term on the right-hand side of \bref{BRST zeta*}
is required for the nilpotency $\delta_B^2 \zeta^*=0$.
We find that the key relation for the nilpotency
\begin{align}
-\pa\cdot \CR
+
\frac{1}{D+2n-4}
(\gamma\pa\cdot\CRsl
+\eta \pa\cdot\CR')
=0
\label{eqn: fer nilpotency}
\end{align}
follows from the triple $\gamma$-traceless condition \bref{psisl'=0}.
Since $\delta_B \zeta^*=(\zeta^*,S^0)$,
the second term on the right-hand side of \bref{BRST zeta*}
requires an additional term
\begin{align}
\frac{n}{D+2n-4}\int \d^Dx\,\left(
\bar\zeta(\gamma\pa\cdot\psisl^*+\eta\pa\cdot\psis')
-\overline{(\gamma\pa\cdot\psisl^*+\eta\pa\cdot\psis{}')}\zeta
\right)
\end{align}
in the action $S^0$.
However this term disappears due to $\zetasl=0$\footnote{$\xi'=0$ follows from $\zetasl=0$.},
and leaves the action \bref{eqn:S^0} unchanged.

\subsection{BRST deformation scheme}\label{sec:BRST deformation}

BRST deformation scheme is very useful in constructing vertices systematically
\cite{BH9304}.
We will explain relevant aspects.

Suppose that
$S$ is a deformation of $S^0$ expanded in a deformation parameter $g$,
\begin{align}
S=S^0+gS^1+g^2S^2+\cdots~.
\end{align}
When $S$ solves the master equation $(S,S)=0$,
$S$ is invariant under the BRST transformation
generated by $S$: $\delta_B^g
 S=(S,S)=0$.
 Here we add the letter $g$ to $\delta_B$
 in order to distinguish it from $\delta_B$ generated by $S^0$
 considered in the previous subsections.
The master equation $(S,S)=0$
means, at the order of $g^n$,
\begin{align}
\sum_{k=0}^n(S^k,S^{n-k})=0.
\label{eqn:master equation g^n}
\end{align}
The equation for $n=0$, $(S^0,S^0)=0$,
is satisfied by definition.
Note that $S^n$ is determined
by the set $\{S^0, S^1,\cdots, S^{n-1}\}
$.

We expand $\delta_B$ 
as
\begin{align}
\delta_B=\Delta+\Gamma,
\end{align}
where
$\Delta$ reduces $agh$ by one while $\Gamma$ leaves it unchanged. 
These $\Delta $ and $\Gamma$
act on fields and antifields as summarized in Table \ref{table:Delta and Gamma}.
\begin{table}[htb]
\begin{align*}
\begin{array}{ccc}\hline 
Z & \Delta(Z) & \Gamma(Z) \\\hline 
\phi_{\mu_1\cdots\mu_{n}} & 0 & \pa_{(\mu_1}c_{\mu_2\cdots\mu_{n})} \\
c_{\mu_2\cdots\mu_{n}} & 0 & 0  \\
\phi^*_{\mu_1\cdots\mu_{n}} & -G_{\mu_1\cdots\mu_{n}}(\phi) & 0  \\
\disp 
c^*_{\mu_2\cdots\mu_{n}} 
&
 -{n}\pa\cdot\phi^*_{\mu_2\cdots\mu_{n}} 
+\frac{{n}}{D+2{n}-6}
\eta_{(\mu_2\mu_3}\pa\cdot \phi^*{}'_{\mu_4\cdots\mu_n)}
& 0  \\\hline 
\psi_{\mu_1\cdots\mu_n} & 0 & \pa_{(\mu_1}\zeta_{\mu_2\cdots\mu_n)} \\
\zeta_{\mu_2\cdots\mu_{n}} & 0 & 0  \\
\psi^*_{\mu_1\cdots\mu_n} & \CR_{\mu_1\cdots\mu_n} & 0  \\
\zeta^*_{\mu_2\cdots\mu_{n}} &
 -n\pa\cdot\psi^*_{\mu_2\cdots\mu_n} 
+\frac{n}{D+2n-4}(\gamma_{(\mu_2}\pa\cdot \psisl^*_{\mu_3\cdots\mu_n)}+\eta_{(\mu_2\mu_3}\pa\cdot \psi^*{}'_{\mu_4\cdots\mu_n)})
& 0  \\\hline 
\end{array}
\end{align*}
 \caption{Action of $\Delta$ and $\Gamma$}
 \label{table:Delta and Gamma}
\end{table}
Correspondingly,
we expand $S^n$
with respect to $agh$.
In this paper, 
we choose a certain cubic vertex as $S^1$,
and find solutions of the master equation $(S,S)=0$
in which $S^n$ is expanded 
as
\begin{align}
 S^n=&\alpha_2^n+\alpha_1^n+\alpha_0^n
\label{eqn:Sn}
\end{align}
where $agh(\alpha^n_i)=i$.
The master equation \bref{eqn:master equation g^n}
at the order of $g^n$
reduces to
the following three equations with respect to $agh$
\begin{align}
&\Gamma \alpha^n_2+\frac{1}{2}\sum_{k=2}^{n-1}(\alpha_2^k,\alpha_1^{n-k})
+\frac{1}{2}\sum_{k=1}^{n-2}(\alpha_1^k,\alpha_2^{n-k})=0
,
\label{eqn:ME Sn 2}\\
&
\Delta \alpha^n_2+\Gamma \alpha^n_1
+\frac{1}{2}\sum_{k=2}^{n-1}(\alpha_2^k,\alpha_0^{n-k})
+\frac{1}{2}\sum_{k=1}^{n-1}(\alpha_1^k,\alpha_1^{n-k})
+\frac{1}{2}\sum_{k=1}^{n-2}(\alpha_0^k,\alpha_2^{n-k})
=0
,
\label{eqn:ME Sn 1}\\
&
\Delta \alpha^n_1+\Gamma \alpha^n_0
+\frac{1}{2}\sum_{k=1}^{n-1}(\alpha_1^k,\alpha_0^{n-k})
+\frac{1}{2}\sum_{k=1}^{n-1}(\alpha_0^k,\alpha_1^{n-k})
=0
.
\label{eqn:ME Sn 0}
\end{align}


\section{Higher-spin gauge model of three bosons}
\label{sec:three bosonic gauge firleds}

We present a higher-spin gauge model of three bosonic tensors.
This is a slight generalization of the model given in \cite{SS2011}.
We give a brief derivation here to make this paper self-contained.

We introduce three gauge fields,
a pair of rank-$n_I$ tensors $\phi_I$ ($I=1,2$)
and a rank-$n$ tensor $\phi$.
As explained in section \ref{sec:free bos},
we introduce ghosts, $c_I$ and $c$, corresponding to the gauge parameters,
and antifields $\{\phi^*_I,\phi^*,c^*_I,c^*\}$ as well.
The free action is 
\begin{align}
S^0=
\intd
\Bigg[&
\sum_{I=1,2}\left(\frac{1}{2}\phi_I^{\mu_1\cdots\mu_{n_I}}G_{\mu_1\cdots\mu_{n_I}}(\phi_I)
+\phi^*_I{}^{\mu_1\cdots \mu_{n_I}}\pa_{(\mu_1}c_I{}_{\mu_2\cdots \mu_{n_I})}
\right)
\nn\\&
+\frac{1}{2}\phi^{\mu_1\cdots\mu_{n}}G_{\mu_1\cdots\mu_{n}}(\phi)
+\phi^*{}^{\mu_1\cdots \mu_{n}}\pa_{(\mu_1}c_{\mu_2\cdots \mu_{n})}
\Bigg]
\label{eqn:action free 3phi}
\end{align}
where $G$ is defined in \bref{eqn:G}.
We derive vertices by applying the BRST deformation scheme.
We expand $S^1$ with respect to $agh$
as $S^1=\alpha_2^1+\alpha_1^1+\alpha^1_0$.
As shown in \cite{SS2011}, 
we may set  $\alpha^1_2=0$.
This is because the $\alpha^1_2$ with $agh=pgh=2$
must be $\Gamma$-exact,
and so it leads to a BRST-exact $S^1$.
The master equation at the order of $g^1$, $(S^0,S^1)=0$, is
\begin{align}
\Gamma \alpha^1_1=&0,
\label{eqn:master a1}\\
\Delta \alpha^1_1+\Gamma \alpha^1_0=&0
\label{eqn:master a0}.
\end{align}
We shall choose $a^1_1$ as
\begin{align}
\alpha^1_1=\intd s^{IJ}\tr
G(\phi_I)^T \pat c\phi^*_J
.
\end{align}
For notational simplicity,
we used a matrix notation.
The integrand means 
\begin{align}
G^{\rho_1\cdots\rho_r\mu_1\cdots\mu_p}(\phi_1)
\pa_{(\mu_1} c_{\mu_2\cdots\mu_p)\nu_1\cdots\nu_q}
\phi^*_2{}^{\nu_1\cdots\nu_q}{}_{\rho_1\cdots\rho_r}
+G^{\rho_1\cdots\rho_r\nu_1\cdots\nu_q}(\phi_2)
\pa_{(\nu_1} c_{\nu_2\cdots\nu_p)\mu_1\cdots\mu_q}
\phi^*_1{}^{\mu_1\cdots\mu_p}{}_{\rho_1\cdots\rho_r}.
\end{align}
Here $p+r=n_1$, $q+r=n_2$ and $p+q=n$.
The $\phi_1$ and $\phi^*_1$ denote $d(p)\times d(r)$ matrices,
while the  $\phi_2$ and $\phi^*_2$ denote $d(q)\times d(r)$ matrices.
When $r=0$, the solution presented here reduces to the one given in \cite{SS2011}.
It is obvious that $\Gamma \alpha^1_1=0$ because of the gauge invariance of $G$.
Acting $-\Delta $ on $\alpha^1_1$, we derive
\begin{align}
-\Delta \alpha^1_1=\intd s^{IJ}\tr G(\phi_I)^T \pat c G(\phi_J)
=\Gamma \intd s^{IJ}\frac{1}{2}\tr G(\phi_I)^T \phi G(\phi_J),
\end{align}
which implies that
 \begin{align}
\alpha^1_0= \intd s^{IJ}\frac{1}{2}\tr G(\phi_I)^T \phi G(\phi_J).
\end{align}
We shall show that
the solution at the order of $g^n$ is
\begin{align}
S^n=\alpha_2^n+\alpha_1^n+\alpha_0^n
\label{eqn:Sn bosonic}
\end{align}
with
\begin{align}
\alpha_2^n=&\intd (s^n)^{IJ}\frac{1}{2}\tr (\pat c \phi_I^*)^T \Phi^{n-2}[\Gt(\pat c \phi^*_J)],
\label{eqn:Sn alpha2 bos}\\
\alpha_1^n=&
\intd  (s^n)^{IJ} \tr (\pat c \phi_I^*)^T \Phi^{n-1}[\Gt(\phi_J)],
\label{eqn:Sn alpha1 bos}\\
\alpha_0^n=&
\intd  (s^n)^{IJ} \frac{1}{2}\tr \phi_I^T \Phi^{n}[\Gt(\phi_J)],
\label{eqn:Sn alpha0 bos}
\end{align}
where $\Phi$ is introduced as $\Phi^0[A]=A$, $\Phi[A]=\Gt(\phi A)$, $\Phi^2[A]=\Gt(\phi \Gt(\phi A))$ 
and so on.
Now, suppose that $\{S^1,S^2,\cdots,S^{n-1} \}$
solves the master equation at the order of $g^k$ ($k=1,\cdots,n-1$).
We will
show that the $\alpha_2^n$, $\alpha_1^n$ and $\alpha_0^n$
coincide with
\bref{eqn:Sn alpha2 bos}, \bref{eqn:Sn alpha1 bos} and \bref{eqn:Sn alpha0 bos}.

First we will derive $\alpha_2^n$ from \bref{eqn:ME Sn 2}.
Because the equation
\begin{align}
-\frac{1}{2}\sum_{k=2}^{n-1}(\alpha_2^k,\alpha_1^{n-k})
-\frac{1}{2}\sum_{k=1}^{n-2}(\alpha_1^k,\alpha_2^{n-k})
=&
\Gamma \intd (s^n)^{IJ} \tr    \frac{1}{2}(\pat c\phi_I^*)^T\Phi^{n-2}[\Gt(\pat c \phi_J^*)]
\end{align}
follows,
$\alpha_2^n$ may be given as \bref{eqn:Sn alpha2 bos}.
Next we will derive $\alpha_1^n$ from \bref{eqn:ME Sn 1}.
We find that
\begin{align}
&
-\Delta \alpha_2^n
-\frac{1}{2} \sum_{k=2}^{n-1}(\alpha_2^k,\alpha_0^{n-k})
-\frac{1}{2}\sum_{k=1}^{n-1}(\alpha_1^k,\alpha_1^{n-k})
-\frac{1}{2} \sum_{k=1}^{n-2}(\alpha_0^k,\alpha_2^{n-k})
\nn\\&=
\Gamma \intd (s^n)^{IJ} \tr \phi_I^T \Phi^{n-1}[\Gt (\pat c \phi^*_J)],
\end{align}
where a useful relation
$\intd A^T\Phi^m[\Gt(B)]=\intd \Phi^m[\Gt(A)]^TB$ is used.
This implies that $\alpha_1^n$ is given as \bref{eqn:Sn alpha1 bos}.
Finally, we derive $\alpha_0^n$ from \bref{eqn:ME Sn 0}.
It is straightforward to obtain
\begin{align}
&
-\Delta \alpha^n_1
-\frac{1}{2}\sum_{k=1}^{n-1}(\alpha_1^k,\alpha_0^{n-k})
-\frac{1}{2}\sum_{k=1}^{n-1}(\alpha_0^k,\alpha_1^{n-k})
=
\Gamma \intd (s^n)^{IJ} \tr \frac{1}{2}\phi_I^T\Phi^{n}[G(\phi_J)]\,.
\end{align}
This implies that 
$\alpha_0^n$ is given as \bref{eqn:Sn alpha0 bos}.
Summarizing the above results,
we have shown that $S^n$ is definitely given as \bref{eqn:Sn bosonic} with \bref{eqn:Sn alpha2 bos}-\bref{eqn:Sn alpha0 bos}.

\medskip

We have derived $S^k$ ($k=1,2,\cdots$).
The free action is given in \bref{eqn:action free 3phi}.
Gathering these together we obtain
the total action
\begin{align}
S=&S^0+\sum_{k=1}^\infty g^k S^k
\nn\\=&
\intd \tr \Bigg[
\frac{1}{2}\phi_I^TG(\phi_I)+
\phi^*_I{}^T\pa c_I
+\frac{1}{2}\phi^TG(\phi)+
\phi^*{}^T\pa c
+\frac{1}{2}\phi_I^T\sum_{k=1}^\infty g^k(s^k)^{IJ}\Phi^k[G(\phi_J)]
\nn\\&\hspace{13mm}
+\phi_I^T\sum_{k=1}^\infty g^k(s^k)^{IJ} \Phi^{k-1}[\Gt(\pat c \phi_J^*)]
+\frac{1}{2}(\pat c\phi_I^*)^T\sum_{k=2}^\infty g^k (s^k)^{IJ} \Phi^{k-2}[\Gt(\pat c \phi_J^*)]
\Bigg].
\label{eqn:infinite bosonic}
\end{align}
We find that the action turns into the form
expanded in $agh$
as
\begin{align}
S=&S_0+S_1+S_2\,,\\
S_0=&\intd\tr\Big[
\frac{1}{2} \phi_I^T
\sum_{k=0}^\infty[(gs)^k]^{IJ}\Phi^k
[G(\phi_J)]
+\frac{1}{2}\phi^TG(\phi)
\Big],
\label{eqn:bos 3 S0}\\
S_1=&\intd \tr \Big[
\phi_I^*{}^T\pa c_I
+\phi^*{}^T\pa c
+\phi_I^T 
\sum_{k=0}^\infty[(gs)^{k+1}]^{IJ}\Phi^k
[\Gt(\pat c \phi_J^*)]
\Big],\\
S_2=&\intd \tr
\frac{1}{2}(\pat c\phi_I^*)^T
\sum_{k=0}^\infty[(gs)^{k+2}]^{IJ}\Phi^k
[\Gt(\pat c \phi_J^*)],
\end{align}
where $agh(S_i)=i$.

\medskip
We have obtained BRST-invariant deformed action of gauge fields
by using the BRST-antifield formalism.
Here we examine the gauge-invariant action $S_0$ in \bref{eqn:bos 3 S0}.
First of all we derive equations of motion.
Varying $S_0$ with respect to $\phi$ and $\phi_I$, we obtain
\begin{align}
&G(\phi)-\frac{1}{2}gs^{IJ}\sum_{k=0}^\infty [(gs)^k]^{IK}\Phi^k[G(\phi_K)]
\left(\sum_{l=0}^\infty[(gs)^l]^{JL}\Phi^l[G(\phi_L)]\right)^T=0,
\\&
\sum_{k=0}^\infty [(gs)^k]^{IJ}\Phi^k[G(\phi_J)]=0.
\label{eqn:eom boson}
\end{align}
Substituting \bref{eqn:eom boson} into the first equation leads to $G(\phi)=0$.
Assuming $|gs\Phi|<1$, 
we find that the second equation of motion \bref{eqn:eom boson} is expressed as 
$(\frac{1}{1-gs\Phi})^{IJ}G(\phi_J)=0$. 
This becomes formally $G(\phi_J)=0$ by acting $(1-gs\Phi)^{KI}$. 
Summarising, we find that the action $S_0$ is free on-shell.
This implies that the gauge model in \cite{SS2011} is also free on-shell, 
since it is a special case of the present gauge model.

Here we examine the gauge invariance of the action $S_0$ in \bref{eqn:bos 3 S0}.
The gauge transformation can be read off from the BRST transformation.
We find that
the gauge transformation with a rank-$(n_I-1)$ tensor parameter $\xi_I$
reamins unchanged
\begin{align}
\delta \phi=0,
~~~
\delta \phi_I=\pa \xi_I,
\end{align}
while the gauge transformation with  a rank-$(n-1)$ tensor parameter $\xi$ 
turns to
\begin{align}
\delta \phi=\pa\xi,
~~~
\delta \phi_I=-
\pat \xi^T
g\sum_{k=0}^\infty[(gs)^k]^{IJ}\Phi^k
[G(\phi_J)].
\end{align}
We note that the gauge algebra is abelian on-shell.

\section{Higher-spin gauge model of two fermions and one boson}\label{sec:three fields}

In the following two sections,
we generalize the gauge model to include fermions.
We examine the on-shell conditions to find that these are free on-shell.

We introduce three gauge fields,
a  rank-$n_1$ tensor-spinor $\psi_1$, a  rank-$n_2$ tensor-spinor $\psi_2$,
and  a rank-$n$ tensor $\phi$.
In addition, 
we introduce corresponding ghosts, $\zeta_I$ $(I=1,2)$ and $c$,
and antifields $\{\psi^*_I,\phi^*,\zeta_I^*,c^*\}$ as well.
As explained in section \ref{sec:free}, the free action for them is given as
\begin{align}
S^0[\Phi,\Phi^*,\Psi,\Psi^*]=&\intd\bigg[ \frac{1}{2}\phi^{\mu_1\cdots\mu_n}G_{\mu_1\cdots\mu_n}
+\phis^{\mu_1\cdots \mu_n}\pa_{(\mu_1}c_{\mu_2\cdots \mu_n)}
\nn\\
&+\sum_{I=1,2}\bigg(\frac{1}{2}\left(
\bar \CR_{\mu_1\cdots\mu_{n_I}}(\psi_I)\psi_I^{\mu_1\cdots\mu_{n_I}}
-
\bar\psi_I^{\mu_1\cdots\mu_{n_I}}\CR_{\mu_1\cdots\mu_{n_I}}(\psi_I)
\right)
\nn\\&\hspace{16mm}
+
\bar\psi^*_I{}^{\mu_1\cdots\mu_{n_I}}\pa_{(\mu_1}\zeta_I{}_{\mu_2\cdots\mu_{n_I})}
-\pa_{(\mu_1}\bar\zeta_I{}_{\mu_2\cdots\mu_{n_I})}\psi^*_I{}^{\mu_1\cdots\mu_{n_I}}
\bigg)
\bigg]\,.
\label{eqn:free action fer 3}
\end{align}

First we consider $S^1=\alpha^1_2+\alpha_1^1+\alpha_0^1$.
We choose a cubic vertex as $S^1$.
As explained in section 3, we may set $\alpha^1_2=0$.
The master equation at the order of $g^1$, $(S^0,S^1)=0$, 
is decomposed with respect to $agh$ into
\bref{eqn:master a1} and \bref{eqn:master a0}.
We shall choose, as $\alpha^1_1$,\begin{align}
\alpha^1_1=i\intd s^{IJ}\tr\bigg[
\bar\psi^*_I\pat c \CR(\psi_J)
+\bar \CR(\psi_I)\pat c^T \psi^*_J
\bigg]
\end{align}
 where $s^{12}=s^{21}=1$ and $s^{11}=s^{22}=0$.
For notational simplicity,
we used the matrix notation.
The first term of the integrand means
\begin{align}
i\bar\psi^*_1{}^{\rho_1\cdots\rho_r\mu_1\cdots\mu_p}
\pa_{(\mu_1} c_{\mu_2\cdots\mu_p)\nu_1\cdots\nu_q} 
\CR^{\nu_1\cdots\nu_q}{}_{\rho_1\cdots\rho_r}(\psi_2)
+
i\bar\psi^*_2{}^{\rho_1\cdots\rho_r\nu_1\cdots\nu_q}
\pa_{(\nu_1} c_{\mu_2\cdots\nu_q)\mu_1\cdots\mu_p} 
\CR^{\mu_1\cdots\mu_p}{}_{\rho_1\cdots\rho_r}(\psi_1),
\end{align}
while the second term means
\begin{align}
i\bar\CR^{\rho_1\cdots\rho_r\mu_1\cdots\mu_p}(\psi_1)
\pa_{(\nu_1} c_{\nu_2\cdots\nu_q)\mu_1\cdots\mu_p} 
\psi^*_2{}^{\nu_1\cdots\nu_q}{}_{\rho_1\cdots\rho_r}
+
i\bar\CR^{\rho_1\cdots\rho_r\nu_1\cdots\nu_q}(\psi_2)
\pa_{(\mu_1} c_{\mu_2\cdots\mu_p)\nu_1\cdots\nu_q} 
\psi^*_1{}^{\mu_1\cdots\mu_p}{}_{\rho_1\cdots\rho_r}.
\end{align}
Here $p+r=n_1$, $q+r=n_2$ and $p+q=n$.
In this notation,
$\psi_1$ and $\psi_1^*$ are $d(p)\times d(r)$ matrices,
$\psi_2$ and $\psi_2^*$ are $d(q)\times d(r)$ matrices,
where $d(s)=\frac{(D-1+s)!}{(D-1)!s!}$.
The index of the derivative in $\pat c$ is always contracted with
one of the indices of the field sitting to its immediate left.

It is obvious that $\Gamma \alpha^1_1=0$ because of the gauge invariance of $\CR$.
Acting $-\Delta $ on $\alpha^1_1$, we derive
\begin{align}
-\Delta \alpha^1_1=-i\intd s^{IJ}\tr
\left[\bar\CR(\psi_I)\pat c \CR(\psi_J)
+\bar \CR(\psi_I)\pat c^T \CR(\psi_J)
\right]
=\Gamma i\intd s^{IJ}\tr \bar \CR(\psi_I)\phi \CR(\psi_J)
.
\end{align}
 As a result, we find
 \begin{align}
\alpha^1_0= i\intd s^{IJ}\tr \bar \CR(\psi_I)\phi \CR(\psi_J)
.
\end{align}

\medskip

We shall  show that the solution of the master equation at the order of $g^n$ is
\begin{align}
S^n=&\alpha^n_2+\alpha^n_1+\alpha^n_0
,
\label{eqn:Sn fer}\\
\alpha^n_2=&\intd i^n (s^n)^{IJ}\tr \bar\psi^*_I\pat c\Psi^{n-2}[\CRt(\pat c^T \psi^*_J)]
,
\label{eqn:Sn alpha2 fer}\\
\alpha^n_1=&\intd  (s^n)^{IJ}\tr\left[
 i^n\bar\psi^*_I\pat c \Psi^{n-1}[\CR(\psi_J)]
 -( -i)^n\overline{\Psi^{n-1}[\CR(\psi_I)]}\pat c^T \psi^*_J
\right]
,
\label{eqn:Sn alpha1 fer}\\
\alpha_0^n=&-\intd i^n(s^n)^{IJ}\tr
\bar\psi_I \Psi^n[\CR(\psi_J)]
.
\label{eqn:Sn alpha0 fer}
\end{align}
Here we have defined $\Psi$ by $\Psi^0[A]=A$, $\Psi[A]=\CRt(\phi A)$, $\Psi^2[A]=\CRt(\phi \CRt(\phi A))$,
and so on.
A useful relation we frequently use is
\begin{align}
\intd \bar A \Psi^k[\CRt(B)]
=(-1)^{k+1}\intd \overline{\Psi^k[\CRt(A)]}B
.
\end{align}
Now, suppose that $\{S^1,S^2,\ldots,S^{n-1} \}$
solves the master equation at the order of $g^k$ ($k=1,\ldots,n-1$).
We will derive $S^n$ expanded as \bref{eqn:Sn fer}
and show that the $\alpha_2^n$, $\alpha_1^n$ and $\alpha_0^n$
coincide with
those given in 
\bref{eqn:Sn alpha2 fer}, \bref{eqn:Sn alpha1 fer} and \bref{eqn:Sn alpha0 fer},
respectively.

The master equation \bref{eqn:master equation g^n}
at the order of $g^n$
reduces to \bref{eqn:ME Sn 2}, \bref{eqn:ME Sn 1} and \bref{eqn:ME Sn 0}.
First we will derive $\alpha_2^n$ from \bref{eqn:ME Sn 2}.
Noting that
\begin{align*}
(\alpha_1^k,\alpha_2^{n-k})=
i^n(s^n)^{IJ}\intd \tr\Big(&
(-1)^{k}\overline{\Psi^{k-1}[\CRt(\pat c^T \psi^*_I)]}
\pat c
\Psi^{n-k-2}[\CRt(\pat c^T \psi^*_J)]
\nn\\
&+
(-1)^{n-k-1}\overline{\Psi^{n-k-2}[\CRt(\pat c^T \psi^*_I)]}
\pat c^T
\Psi^{k-1}[\CRt(\pat c^T \psi^*_J)]
\Big)
,
\\
(\alpha_2^k,\alpha_1^{n-k})=
i^n(s^n)^{IJ}\intd \tr\Big(&
(-1)^{n-k}\overline{\Psi^{n-k-1}[\CRt(\pat c^T \psi^*_I)]}
\pat c
\Psi^{k-2}[\CRt(\pat c^T \psi^*_J)]
\nn\\
&+
(-1)^{k-1}\overline{\Psi^{k-2}[\CRt(\pat c^T \psi^*_I)]}
\pat c^T
\Psi^{n-k-1}[\CRt(\pat c^T \psi^*_J)]
\Big)
,
\end{align*}
we obtain
\begin{align}
&-\frac{1}{2}\sum_{k=1}^{n-2}(\alpha_1^k,\alpha_2^{n-k})
-\frac{1}{2}\sum_{k=2}^{n-1}(\alpha_2^k,\alpha_1^{n-k})
\nn\\
&=\intd i^n(s^n)^{IJ} \tr\sum_{l=0}^{n-3}
(-1)^l\overline{\Psi^l[\CRt(\pat c^T\psi^*_I)]}
\pa c\Psi^{n-3-l}[\CRt(\pat c^T\psi_J^*)]
\nn\\
&=
\Gamma\intd  i^n (s^n)^{IJ} \tr \bar \psi^*_I\pat c\Psi^{n-2}[\CRt(\pat c^T \psi_J^*)]
.
\end{align}
This implies that $\alpha_2^n$ is given as \bref{eqn:Sn alpha2 fer}.

Next we will derive $\alpha_1^n$ from \bref{eqn:ME Sn 1}.
Using \bref{eqn:Sn alpha2 fer}, we derive
\begin{align}
&
-\Delta \alpha_2^n
-\frac{1}{2} \sum_{k=2}^{n-1}(\alpha_2^k,\alpha_0^{n-k})
-\frac{1}{2} \sum_{k=1}^{n-2}(\alpha_0^k,\alpha_2^{n-k})
\nn\\
&=
 i^n (s^n)^{IJ}\intd \tr\bigg(
 \sum_{k=0}^{n-2}
(-1)^{k+1}\overline{\Psi^k[\CR(\psi_I)]}\pat c \Psi^{n-k-2}[\CRt(\pat c^T \psi^*_J)]
\nn\\&\hspace{37mm}
+\sum_{l=0}^{n-2}(-1)^l\overline{\Psi^l[\CRt(\pat c^T \psi_I^*)]}\pat c^T\Psi^{n-2-l}[\CR(\psi_J)]
\bigg)
.
\end{align}
On the other hand, one finds
\begin{align}
-\frac{1}{2}\sum_{k=1}^{n-1}(\alpha_1^k,\alpha_1^{n-k})
=&
i^n  \sum_{k=1}^{n-1}
(s^n)^{IJ}\intd \tr\bigg(
(-1)^{k}\overline{\Psi^{k-1}[\CR(\psi_I)]} \pat c^T\Psi^{n-k-1}[\CRt(\pat c^T\psi^*_J)]
\nn\\&\hspace{37mm}
+(-1)^{k+1}\overline{\Psi^{k-1}[\CRt(\pat c^T \psi^*_I)]}\pat c \Psi^{n-k-1}[\CR(\psi_J)]
\bigg)
.
\end{align}
Combining these results together
we obtain
\begin{align}
&
-\Delta \alpha_2^n
-\frac{1}{2} \sum_{k=2}^{n-1}(\alpha_2^k,\alpha_0^{n-k})
-\frac{1}{2}\sum_{k=1}^{n-1}(\alpha_1^k,\alpha_1^{n-k})
-\frac{1}{2} \sum_{k=1}^{n-2}(\alpha_0^k,\alpha_2^{n-k})
\nn\\&=
 i^n (s^n)^{IJ}\intd \tr\bigg(
 \sum_{k=0}^{n-2}
(-1)^{k+1}\overline{\Psi^k[\CR(\psi_I)]}\pa c \Psi^{n-k-2}[\CRt(\pat c^T \psi^*_J)]
\nn\\&\hspace{37mm}
+\sum_{l=0}^{n-2}(-1)^l\overline{\Psi^l[\CRt(\pat c^T \psi_I^*)]} \pa c\Psi^{n-2-l}[\CR(\psi_J)]
\bigg)
\nn\\&=
\Gamma \intd i^n (s^n)^{IJ}\tr \Big(
\bar\psi^*_I\pat c \Psi^{n-1}[\CR(\psi_J)]
-(-1)^n
\overline{\Psi^{n-1}[\CR(\psi_I)]}\pat c^T \psi^*_J
\Big)
.
\end{align}
This implies that $\alpha_1^n$ is given as in \bref{eqn:Sn alpha1 fer}.

Finally, we solve \bref{eqn:ME Sn 0} for $\alpha_0^n$.
It is straightforward to derive
\begin{align}
&
-\Delta \alpha^n_1
-\frac{1}{2}\sum_{k=1}^{n-1}(\alpha_1^k,\alpha_0^{n-k})
-\frac{1}{2}\sum_{k=1}^{n-1}(\alpha_0^k,\alpha_1^{n-k})
\nn\\&=
i^n (s^n)^{IJ}\intd\tr
\bigg(
\sum_{k=0}^{n-1}
(-1)^{k+1}\overline{\Psi^k[\CR(\psi_I)]}\pat c \Psi^{n-k-1}[\CR(\psi_J)]
\nn\\&\hspace{37mm}
+
\sum_{k=1}^{n}
(-1)^k\overline{\Psi^{k-1}[\CR(\psi_I)]}\pat c^T \Psi^{n-k}[\CR(\psi_J)]
\bigg)
\nn\\&=
i^n (s^n)^{IJ}\intd\tr
\sum_{k=0}^{n-1}
(-1)^{k+1}\overline{\Psi^k[\CR(\psi_I)]}\pa c \Psi^{n-k-1}[\CR(\psi_J)]
\nn\\&=
\Gamma \intd (-i^n)(s^n)^{IJ}\tr
\bar\psi_I \Psi^n[\CR(\psi_J)]
.
\end{align}
This implies that 
$\alpha_0^n$ is given as in \bref{eqn:Sn alpha0 fer}.

Summarizing the above results,
we have shown that $S^n$ is definitely given in \bref{eqn:Sn fer}-\bref{eqn:Sn alpha0 fer}.

\subsection{BRST-invariant deformed action of three gauge fields}

In the above we have derived $S^k$ ($k=1,2,\cdots$).
The free action $S^0$ is
given in \bref{eqn:free action fer 3}.
Combining these together we obtain the total action
\begin{align}
S=&S^0
+\sum_{k=1}^\infty g^k S^k
\nn\\=&
\intd \tr\bigg[
\frac{1}{2}\phi^TG(\phi)+
\phi^*{}^T\pa c
+\frac{1}{2}\left(\bar\CR(\psi_I)\psi_I-\bar\psi_I\CR(\psi_I)\right)
+\bar\psi^*_I \pa\zeta_I
-\pa\bar\zeta_I \psi^*_I
\nn\\&\hspace{13mm}
-\bar\psi_I\sum_{k=1}^\infty [(igs)^k]^{IJ}\Psi^k[\CR(\psi_J)]
+\bar\psi^*_I \pat c\sum_{k=1}^\infty [(igs)^k]^{IJ} \Psi^{k-1}[\CR(\psi_J)]
\nn\\&\hspace{13mm}
-\sum_{k=1}^\infty [(-igs)^k]^{IJ}\overline{\Psi^{k-1}[\CR(\psi_I)]}\pat c^T \psi^*_J
+\bar\psi^*_I\pat c\sum_{k=2}^\infty [(igs)^k]^{IJ} \Psi^{k-2}[\CRt(\pat c^T\psi^*_J)]
\bigg].
\label{eqn:S infinite}
\end{align}
This action contains terms of all orders  in the deformation parameter $g$
and is BRST-invariant exactly.
We note that the BRST-invariance of the action is ensured even if $g$ is not small.

We find that the action $S$ turns to the form
expanded in $agh$
as
\begin{align}
S=&S_0+S_1+S_2,
\label{eqn:S 3 agh AdS}\\
S_0=&\intd\tr\bigg[
\frac{1}{2}\phi^TG(\phi)
-\bar\psi_I
\sum_{k=0}^\infty[(igs)^k]^{IJ}\Psi^k[\CR(\psi_J)]
\bigg],
\label{eqn:S 3 agh S0 AdS}\\
S_1=&\intd\tr\bigg[
\phi^*{}^T\pa c
+\bar\psi^*_I \pa\zeta_I
-\pa\bar\zeta_I \psi^*_I
\nn\\&\hspace{13mm}
+\bar\psi^*_I\pat c
\sum_{k=0}^\infty[(igs)^{k+1}]^{IJ}\Psi^k[\CR(\psi_J)]
-
\sum_{k=0}^\infty\overline{[(igs)^{k+1}]^{IJ}\Psi^k
[\CR(\psi_J)]
}\pat c^T\psi^*_I
\bigg],\\
S_2=&\intd\tr
\bar\psi^*_I \pat c 
\sum_{k=0}^\infty[(igs)^{k+2}]^{IJ}\Psi^k
[\CRt(\pat c^T \psi^*_J)],
\label{eqn:S2 3 agh AdS}
\end{align}
where $agh(S_i)=i$.
This action $S$ contains all orders in $g$,
and invariant under the BRST-transformation $\delta_B^g S=(S,S)=0$.

We have obtained a BRST-invariant deformed action of three gauge fields
by using the BRST-antifield formalism.
Here we examine the gauge-invariant action $S_0$ in \bref{eqn:S 3 agh S0 AdS}.
First of all we derive equations of motion.
Varying $S_0$ with respect to $\phi$ and $\bar\psi_I$, we obtain
\begin{align}
&G(\phi)+igs^{IJ}\sum_{k=0}^\infty [(igs)^k]^{IK}\Psi^k[\CR(\psi_K)]
\sum_{l=0}^\infty \overline{[(igs)^l]^{JL}\Psi^l[\CR(\psi_L)]}=0,
\\&
\sum_{k=0}^\infty [(igs)^k]^{IJ}\Psi^k[\CR(\psi_J)]=0.
\label{eqn:eom fermion}
\end{align}
Substituting \bref{eqn:eom fermion} into the first equation leads to $G(\phi)=0$.
Furthermore,
assuming $|igs\Psi|<1$,
the second equation of motion \bref{eqn:eom fermion} can be expressed as $(\frac{1}{1-igs\Psi})^{IJ}\CR(\psi_J)=0$.
Applying $(1-igs\Psi)^{KI}$ to this, we formally get $\CR(\psi_J)=0$.
As a result, we find that
the model including two fermions and a boson reduce to free on-shell.

The gauge transformation can be read off from the BRST transformation.
We find that
the gauge transformation with a rank-$(n_I-1)$ tensor-spinor parameter $\epsilon_I$
remains unchanged
\begin{align}
\delta \phi=0,~~~
\delta \psi_I=\pa\epsilon_I,
\end{align}
while the gauge transformation with  a rank-$(n-1)$ tensor parameter $\xi$ 
turns into
\begin{align}
\delta \phi=\pa \xi,~~~
\delta \psi_I=\pat\xi
\sum_{k=0}^\infty[(igs)^{k+1}]^{IJ}\Psi^k
[\CR(\psi_J)].
\end{align}
The latter gauge transformation is deformed by vertices.
However we note that the gauge algebra remains abelian on-shell.
In fact, applying the gauge transformation twice on $\psi_I$ leads to terms
which vanish due to \bref{eqn:eom fermion}.

\section{Higher-spin gauge model of a fermion and a boson}\label{sec:two fields}

In this section,
we construct a model which formed by two gauge fields,
a rank-$n_1$ tensor-spinor $\psi$ and a rank-$n$ tensor $\phi$.
This model is found to be free on-shell.
By introducing corresponding ghosts, $\zeta$ and $c$,
and antifields $\{\psi^*,\phi^*,\zeta^*,c^*\}$,
the free action is given as
\begin{align}
S^0[\Phi,\Phi^*]=&\intd\bigg[ \frac{1}{2}\phi^{\mu_1\cdots\mu_n}G_{\mu_1\cdots\mu_n}
+\phis^{\mu_1\cdots \mu_n}\pa_{(\mu_1}c_{\mu_2\cdots \mu_n)}
\nn\\
&+\frac{1}{2}\left(
\bar \CR_{\mu_1\cdots\mu_{n_1}}(\psi)\psi^{\mu_1\cdots\mu_{n_1}}
-
\bar\psi^{\mu_1\cdots\mu_{n_1}}\CR_{\mu_1\cdots\mu_{n_1}}(\psi)
\right)
\nn\\&\hspace{16mm}
+
\bar\psi^*{}^{\mu_1\cdots\mu_{n_1}}\pa_{(\mu_1}\zeta_{\mu_2\cdots\mu_{n_1})}
-\pa_{(\mu_1}\bar\zeta_{\mu_2\cdots\mu_{n_1})}\psi^*{}^{\mu_1\cdots\mu_{n_1}}
\bigg].
\end{align}

Applying the BRST deformation scheme,
we can derive a deformed action in the similar manner explained in the previous section.
We present the result below.
We find that the solution of the master equation \bref{eqn:master equation g^n} at the order of $g^n$ ($n\geq 1$) is
\begin{align}
S^n=&\alpha^n_2+\alpha^n_1+\alpha^n_0,\\
\alpha^n_2=&\intd i^n \tr \bar\psi^*\pat c\Psi^{n-2}[\CRt(\pat c^T \psi^*)],\\
\alpha^n_1=&\intd \tr\left[
 i^n\bar\psi^*\pat c \Psi^{n-1}[\CR(\psi)]
 -( -i)^n\overline{\Psi^{n-1}[\CR(\psi)]}\pat c^T \psi^*
\right],\\
\alpha_0^n=&-\intd i^n\tr
\bar\psi \Psi^n[R(\psi)].
\end{align}
We used the matrix notation in which
$\psi$ and $\psi^*$ are $d(p)\times d(r)$ matrices,
$\bar\psi$ and $\bar\psi^*$ are $d(r)\times d(p)$ matrices
and 
$\phi$ and $\pat c$ are $d(p)\times d(p)$ matrices,
where $p+r=n_1$ and $2p=n$.

We find that the action turns to the form
expanded in $agh$
as
\begin{align}
S=&S_0+S_1+S_2,
\label{eqn:S 2 AdS}\\
S_0=&\intd\tr\bigg[
\frac{1}{2}\phi^TG(\phi)
-\bar\psi
\sum_{k=0}^\infty(ig)^k\Psi^k
[\CR(\psi)]
\bigg],
\label{eqn:S 2 S0 AdS}\\
S_1=&\intd\tr\bigg[
\phi^*{}^T\pa c
+\bar\psi^* \pa\zeta
-\pa\bar\zeta \psi^*
\nn\\&\hspace{13mm}
+\bar\psi^*\pat c 
\sum_{k=0}^\infty(ig)^{k+1}\Psi^k
[\CR(\psi)]
-
\sum_{k=0}^\infty\overline{(ig)^{k+1}\Psi^k
[\CR(\psi)]}\,\pat c^T\psi^*
\bigg],\\
S_2=&\intd\tr
\bar\psi^*\pat c
\sum_{k=0}^\infty(ig)^{k+2}\Psi^k
[\CRt(\pat c^T \psi^*)].
\label{eqn:S2 2 AdS}
\end{align}
This action $S$ contains all orders in $g$,
and invariant under the BRST-transformation $\delta_B^g S=(S,S)=0$.

We have obtained a BRST-invariant deformed action of two gauge fields
by using the BRST-antifield formalism.
Varying $S_0$ with respect to $\phi$ and $\bar\psi$, we obtain equations of motion
\begin{align}
&G(\phi)+ig\sum_{k=0}^\infty (ig)^k\Psi^k[\CR(\psi)]
\sum_{l=0}^\infty \overline{(ig)^l\Psi^l[\CR(\psi)]}=0,
\\&
\sum_{k=0}^\infty (ig)^k\Psi^k[\CR(\psi)]=0.
\label{eqn:eom one fermion}
\end{align}
By the same argument given in section \ref{sec:three fields},
we find  that the model including a fermion and a boson reduces to free on-shell.

Next we comment on the gauge invariance of the action $S_0$ in \bref{eqn:S 2 S0 AdS}.
We find that
the gauge transformation with a rank-$(n_1-1)$ tensor-spinor parameter $\epsilon$
remains unchanged
\begin{align}
\delta \phi=0,~~~
\delta \psi=\pa\epsilon,
\end{align}
while the gauge transformation with  a rank-$(n-1)$ tensor parameter $\xi$ 
changes into
\begin{align}
\delta \phi=\pa \xi,~~~
\delta \psi=\pat\xi
\sum_{k=0}^\infty(ig)^{k+1}\Psi^k
[\CR(\psi)].
\end{align}
The gauge transformation is deformed by the vertex,
but
the gauge algebra remains abelian on-shell.

\section{Higher-spin gauge models on AdS spaces}\label{sec:AdS}
The Higher-spin gauge models obtained in sections
\ref{sec:three bosonic gauge firleds},  \ref{sec:three fields} and \ref{sec:two fields}
are generalized to those on AdS spaces.
For this purpose, we formulate the BRST deformation scheme on AdS spaces.

\subsection{Bosonic gauge fields on AdS spaces}
\label{sec:bos AdS}
We derive a higher-spin gauge model on AdS spaces by applying the BRST deformation scheme built on AdS spaces.
The Fronsdal tensor
for a rank-$n$
tensor gauge field  on AdS spaces
is
\begin{align}
\msF_{\mu_1\cdots\mu_n}(\phi)=&
\square\phi_{\mu_1\cdots\mu_n}
-\nabla_{(\mu_1}\nabla^\sigma\phi_{\mu_2\cdots\mu_n)\sigma}
+\frac{1}{2}\nabla_{(\mu_1}\nabla_{\mu_2}\phi'_{\mu_3\cdots\mu_n)}
-\frac{m^2}{l^2}\phi_{\mu_1\cdots\mu_n}
-\frac{2}{l^2}\gb_{(\mu_1\mu_2}\phi'_{\mu_3\cdots\mu_n)},
\end{align}
where $\square =\gb^{\mu\nu}\nabla_{\mu}\nabla_\nu$
and
$m^2=n^2+n(D-6)-2(D-3)$\,.
$\msF$ reduces to $F$ in  \bref{Fronsdal bos flat} in the flat limit.
The covariant derivative satisfies
\begin{align}
[\nabla_\mu,\nabla_\nu]\phi_{\rho_1\cdots\rho_n}
=R_{\mu\nu(\rho_1}{}^\lambda\phi_{\rho_2\cdots\rho_n)\lambda},~~~
R_{\mu\nu\rho\sigma}=
-\frac{1}{l^2}(\gb_{\mu\rho}\gb_{\nu\sigma}-\gb_{\mu\sigma}\gb_{\nu\rho})\,,
\end{align}
where $l$ denotes the radius of the AdS space. 
Under the gauge transformation
\begin{align}
\delta \phi_{\mu_1\cdots\mu_n}
=\nabla_{(\mu_1}\xi_{\mu_2\cdots\mu_n)},
\label{eqn:gauge trans bos AdS}
\end{align}
the Fronsdal tensor changes to
\begin{align}
\delta \msF=\frac{1}{2}\nabla_{(\mu_1}\nabla_{\mu_2}\nabla_{\mu_3}
\xi'_{\mu_4\cdots\mu_n)}
-\frac{4}{l^2}\gb_{(\mu_1\mu_2}\nabla_{\mu_3}\xi'_{\mu_4\cdots\mu_n)}\,,
\end{align}
which implies $\delta \msF=0$ due to $\xi'=0$.
By introducing a rank-$(n-1)$ tensor ghost $c_{\mu_2\cdots\mu_n}$
and antifields $\{\phi^*,c^*\}$,
the gauge symmetry may be lifted to the BRST symmetry\footnote{We comment on the case 
$D+2n-6=0$, namely $(D,n)=(6,0)$ and $(4,1)$.
When $(D,n)=(6,0)$, gauge transformation \bref{eqn:gauge trans bos AdS}
becomes trivial.
When $(D,n)=(4,1)$,
$\msG=\msF$ as $\msF'=0$.
Since $\nabla\cdot \msG=\nabla\cdot(\square \phi-\nabla\nabla\cdot\phi-\frac{m^2}{l^2}\phi)=0$,
the second term on the right-hand side of the last equation in \bref{eqn:BRST bos AdS}
is not needed.
}
\begin{align}
\delta_B\phi=&\nabla c,&
\delta_B c=&0,\nn\\
\delta_B\phi^*=&-\msG(\phi),&
\delta_B c^*=&-n\nabla\cdot \phi^*
+\frac{n}{D+2n-6}\gb\nabla\cdot \phis',
\label{eqn:BRST bos AdS}
\end{align}
where $\msG$ is defined as
\begin{align}
\msG_{\mu_1\cdots\mu_n}(A)=&\msF_{\mu_1\cdots\mu_n}(A)
-\frac{1}{2}\gb_{(\mu_1\mu_2}\msF'_{\mu_3\cdots\mu_n)}(A)
\label{eqn:G AdS}
\end{align}
where $A$ is a symmetric rank-$n$ tensor.
The last term on the right-hand side of the last equation in \bref{eqn:BRST bos AdS} is required for the nilpotency $\delta_B^2c^*=0$.
A key relation for the nilpotency
\begin{align}
\nabla\cdot \msG -\frac{1}{D+2n-6}\gb\nabla\cdot \msG'=0
\end{align}
follows from the double-traceless condition $\phi''=0$.
The free action $S^0$
which reproduces  the BRST transformation \bref{eqn:BRST bos AdS} by $\delta_B=(X,S^0)$,
is
\begin{align}
S^0=
\intd\sqrt{-\gb}
\Bigg[&
\sum_{I=1,2}\left(\frac{1}{2}\phi_I^{\mu_1\cdots\mu_{n_I}}\msG_{\mu_1\cdots\mu_{n_I}}(\phi_I)
+\phi^*_I{}^{\mu_1\cdots \mu_{n_I}}\nabla_{(\mu_1}c_I{}_{\mu_2\cdots \mu_{n_I})}
\right)
\nn\\&
+\frac{1}{2}\phi^{\mu_1\cdots\mu_{n}}\msG_{\mu_1\cdots\mu_{n}}(\phi)
+\phi^*{}^{\mu_1\cdots \mu_{n}}\nabla_{(\mu_1}c_{\mu_2\cdots \mu_{n})}
\Bigg].
\label{eqn:action free 3phi AdS}
\end{align}

By applying the BRST deformation scheme, we can obtain vertices
in the manner similar to one explained above.
The action of the BRST differentials $\Delta$ and $\Gamma$ is 
summarized in Table \ref{table:Delta and Gamma AdS}.
It is straightforward to see that
\begin{align}
\intd \sqrt{-\gb}\,\tr \varphi_1^T \tilde \msG(\varphi_2)
=\intd \sqrt{-\gb}\,\tr \tilde \msG(\varphi_1)^T \varphi_2
\label{eqn:formula bosonic AdS}
\end{align}
where $\varphi_i$ ($i=1,2$) are totally-symmetric tensors.
We have introduced
\begin{align}
\tilde \msG_{\mu_1\cdots\mu_n}(A)=&
\msG_{\mu_1\cdots\mu_n}(A)
+\frac{1}{4}\gb_{(\mu_1\mu_2}\nabla_{\mu_3}\nabla_{\mu_4} A''_{\mu_5\cdots\mu_n)}.
\end{align}
Using this relation \bref{eqn:formula bosonic AdS},
we obtain the action for three bosonic higher-spin gauge fields
on AdS spaces
\begin{align}
S=&S_0+S_1+S_2,\\
S_0=&\intd\sqrt{-\gb}\,\tr\Big[
\frac{1}{2} \phi_I^T
\sum_{k=0}^\infty[(gs)^k]^{IJ}\varPhi^k
[\msG(\phi_J)]
+\frac{1}{2}\phi^T \msG(\phi)
\Big],
\label{eqn:action AdS 3bos S0}\\
S_1=&\intd\sqrt{-\gb}\, \tr \Big[
\phi_I^*{}^T\nabla c_I
+\phi^*{}^T\nabla c
+\phi_I^T
\sum_{k=0}^\infty[(gs)^{k+1}]^{IJ}\varPhi^k
[\msGt(\tilde\nabla c \phi_J^*)]
\Big],\\
S_2=&\intd\sqrt{-\gb}\, \tr
\frac{1}{2}(\tilde\nabla c\phi_I^*)^T
\sum_{k=0}^\infty[(gs)^{k+2}]^{IJ}\varPhi^k
\msGt(\tilde\nabla c \phi_J^*).
\end{align}
$\varPhi$ is introduced as
$\varPhi^0[A]=A$,
$\varPhi[A]=\msGt(\phi A)$,
$\varPhi^2[A]=\msGt(\phi \msGt(\phi A))$ 
and so on.
Varying $S_0$
with respect to $\phi$ and $\phi_J$
we find that $\msG(\phi)=0$ and $\msG(\phi_J)=0$
assuming $|gs\varPhi|<1$.

The  action  $S_0$ in \bref{eqn:action AdS 3bos S0}
is found to be invariant under 
the gauge transformation with a rank-$(n_I-1)$ tensor parameter $\xi_I$
\begin{align}
\delta \phi=0,
~~~
\delta \phi_I=\nabla \xi_I,
\end{align}
and
the gauge transformation with  a rank-$(n-1)$ tensor parameter $\xi$ 
\begin{align}
\delta \phi=\nabla\xi,
~~~
\delta \phi_I=-
\tilde\nabla \xi^T
g\sum_{k=0}^\infty[(gs)^{k}]^{IJ}\varPhi^k
[\msG(\phi_J)].
\end{align}

\subsection{Fermionic gauge fields on AdS spaces}
\label{sec:fer AdS}

First of all we introduce the covariant derivative on a tensor-spinor
and the Fronsdal tensor on AdS spaces.
The covariant derivative\footnote{
Recall that Lorentz transformation law of a field $\varphi$
is $\delta \varphi=\frac{1}{2}\epsilon ^{ab} S_{ab}\varphi$,
where $\epsilon ^{ab}$ is an antisymmetric parameter
and $a,b$ denote local Lorentz indices.
The $S_{ab}$ is the Lorentz generator
satisfying 
$S_{ab}\phi=0$ for a spin-$0$ $\phi$, $S_{ab}\psi=\frac{1}{2}\gamma_{ab}\psi$ for a spin-$\frac{1}{2}$ $\psi$
and $(S_{ab})_{cd}A^d=(\eta_{ac}\eta_{bd}-\eta_{ad}\eta_{bc})A^d$
for a spin-$1$ $A^a$.
}
of a field $\varphi$ is introduced as
$D_\mu \varphi=\pa_\mu \varphi +\frac{1}{2}\omega^{ab}_\mu S_{ab}\varphi$
where $\omega^{ab}$ is the spin-connection.
The total covariant derivative $\msD_\mu$
acts on an arbitrary rank-$n$ tensor-spinor $\psi_{\mu_1\cdots\mu_n}$
as
\begin{align}
\msD_\mu \psi_{\mu_1\cdots\mu_n}=
D_\mu \psi_{\mu_1\cdots\mu_n}
-\Gamma_{\mu(\mu_1}^\lambda \psi_{\mu_2\cdots\mu_n)\lambda}
=
\nabla_\mu \psi_{\mu_1\cdots\mu_n}
+\frac{1}{4}\omega^{ab}_\mu\gamma_{ab}  \psi_{\mu_1\cdots\mu_n}.
\end{align}
The commutation relation of $\msD_\mu$ on a tensor-spinor $\psi_{\mu_1\cdots\mu_n}$
turns to 
\begin{align}
[\msD_\mu,\msD_\nu] \psi_{\mu_1\cdots\mu_n}
=&
R_{\mu\nu(\mu_1}{}^\rho\psi_{\mu_2\cdots\mu_n)\rho}
+\frac{1}{4}R^{ab}{}_{\mu\nu}\gamma_{ab}\psi_{\mu_1\cdots\mu_n}
\nn\\
=&-\frac{1}{2l^2}\gamma_{\mu\nu}\psi_{\mu_1\cdots\mu_n}
-\frac{1}{l^2}\left(\gb_{\mu(\mu_1}\psi_{\mu_2\cdots\mu_n)\nu}
-\gb_{\nu(\mu_1}\psi_{\mu_2\cdots\mu_n)\mu}
\right)
\end{align}
where $l$ denotes the radius of the AdS space. 
We note that
$\msD_\mu\gamma^\nu \psi=\gamma^\nu\msD_\mu \psi$
\footnote{
This can be shown as
\begin{align}
\msD_\mu \gamma^\nu \psi_{\mu_1\cdots\mu_n}
=&e^\nu_a \msD_\mu( \gamma^a\psi_{\mu_1\cdots\mu_n})
\nn\\=&
e^\nu_a\big(\nabla_\mu(\gamma^a\psi_{\mu_1\cdots\mu_n})
+\omega_\mu^a{}_b\gamma^b \psi_{\mu_1\cdots\mu_n}
+\frac{1}{4}\omega^{cd}_\mu\gamma_{cd} \gamma^a \psi_{\mu_1\cdots\mu_n}
\big)
\nn\\=&
e^\nu_a\gamma^a\big(\nabla_\mu\psi_{\mu_1\cdots\mu_n}
+\frac{1}{4}\omega^{cd}_\mu\gamma_{cd} \psi_{\mu_1\cdots\mu_n}
\big)\nn\\=&
\gamma^\nu\msD_\mu \psi_{\mu_1\cdots\mu_n}
\nn
\end{align}
where $\gamma^\nu=e^\nu_a\gamma^a$ and $\msD_\mu e^\nu_a=0$ are used in the first equality.
},
which is useful in deriving this equation.
In the second equality, we have used the fact that
$R_{\mu\nu\rho\sigma}=-\frac{1}{l^2}(\gb_{\mu\rho}\gb_{\nu\sigma}-\gb_{\mu\sigma}\gb_{\nu\rho})$
for AdS spaces.

Fronsdal tensor for a rank-$n$ tensor-spinor $\psi_{\mu_1\cdots\mu_n}$ on AdS spaces is given as
\cite{ST0311}
\begin{align}
\msS_{\mu_1\cdots\mu_n}=&i\left(
\SDsl\psi_{\mu_1\cdots\mu_n}
-\msD_{(\mu_1}\psisl_{\mu_2\cdots\mu_n)}
+\frac{m}{2l}\psi_{\mu_1\cdots\mu_n}
+\frac{1}{2l}\gamma_{(\mu_1}\psisl_{\mu_2\cdots\mu_n)}
\right)
,
\end{align}
where $m=D+2n-4$.
It reduces to $\CS$ in \bref{eqn:Fronsdal fer flat} in the limit $l\to \infty$.
This is invariant under the gauge transformation
with a rank-$(n-1)$ tensor-spinor gauge parameter $\epsilon$ \cite{ST0311}
\begin{align}
\delta \psi_{\mu_1\cdots\mu_n}
&=\hat\msD_{(\mu_1} \epsilon_{\mu_2\cdots\mu_n)}
\label{eqn:gauge trans fermion AdS}
\end{align}
if $\epsilon$ satisfies the $\gamma$-traceless condition $\epsilonsl=0$,
which is assumed below.
We have introduced $\hat \msD$ by
\begin{align}
\hat\msD_{(\mu_1} \epsilon_{\mu_2\cdots\mu_n)}&\equiv\msD_{(\mu_1} \epsilon_{\mu_2\cdots\mu_n)}
+\frac{1}{2l}\gamma_{(\mu_1}  \epsilon_{\mu_2\cdots\mu_n)}.
\end{align}

The action which leads to the Fronsdal equation
 $\CS=0$ 
is  given by
\begin{align}
S_\psi
&=\int \d ^Dx\sqrt{-\gb} \frac{1}{2}\left[
\bar \msR_{\mu_1\cdots\mu_n}\psi^{\mu_1\cdots\mu_n}
-
\bar\psi^{\mu_1\cdots\mu_n}\msR_{\mu_1\cdots\mu_n}
\right]\,,
\label{eqn:action psi AdS}
\end{align}
where we introduced 
$\msR_{\mu_1\cdots\mu_n}$ as
\begin{align}
\msR_{\mu_1\cdots\mu_n}(\psi)&\equiv
\msS_{\mu_1\cdots\mu_n}
-\frac{1}{2}\gamma_{(\mu_1}\msSsl_{\mu_2\cdots\mu_n)}
-\frac{1}{2}\gb_{(\mu_1\mu_2} \msS'_{\mu_3\cdots\mu_n)}
\,.
\label{eqn:R AdS}
\end{align}
An important relation we frequently use below is
\begin{align}
\intd \sqrt{-\gb}\bar{\tilde\msR}(\psi_1)\psi_2=-\intd \sqrt{-\gb}\bar\psi_1 \msRt(\psi_2)
\end{align}
where
$\psi_i$ are arbitrary rank-$n$ tensor-spinors,
and $\msRt(A)$ is defined by
\begin{align}
\msRt(A)\equiv &\,\msR(A)-\frac{i}{2}\gb\msD\Asl'
\end{align}
where $A$ is arbitrary rank-$n$ tensor-spinor.
Using this relation,
we derive $\msR=0$ from the variation of $S_\psi$.
In the manner similar to one explained
in section \ref{sec:free fer},
$\msR=0$ leads to $\msS=0$.
The action $S_\psi$ is gauge invariant because 
$\msR$ is gauge invariant $\msR(\delta\psi)=0$.

Corresponding to the gauge parameter $\epsilon_{\mu_2\cdots\mu_{n}}$,
we introduce a Grassmann-even tensor-spinor ghost $\zeta_{\mu_2\cdots\mu_{n}}$
which
is
$\gamma$-traceless $\zetasl_{\mu_3\cdots\mu_{n}}=0$.
The gauge invariance of $S_\psi$ is encoded to
the BRST invariance under the BRST transformation
\begin{align}
\delta_B \psi_{\mu_1\cdots\mu_n}&=
\hat\msD_{(\mu_1} \zeta_{\mu_2\cdots\mu_n)},~~~
\delta_B \zeta_{\mu_2\cdots\mu_n}=0~.
\label{BRST psi zeta AdS}
\end{align}
The action $S_\psi$
in \bref{eqn:action psi AdS}
can be extended to $S^0[\Psi,\Psi^*]$ such that
the BRST transformation of a functional $X(\Psi^A,\Psi^*_A)$
is expressed as
$\delta_B X=(X,S^0)$
\,.
In the present case, 
$S^0$ may be given as
\begin{align}
S^0[\Psi,\Psi^*]=
S_\psi
+\int \d ^Dx\sqrt{-\gb} \left(
\bar\psi^*{}^{\mu_1\cdots\mu_n}\hat\msD_{(\mu_1}\zeta_{\mu_2\cdots\mu_n)}
-\overline{\hat\msD_{(\mu_1}\zeta_{\mu_2\cdots\mu_n)}}\psis{}^{\mu_1\cdots\mu_n}
\right),
\label{eqn:S^0 psi AdS}
\end{align}
which leads to
\bref{BRST psi zeta AdS}
and
\begin{align}
\delta_B\psi^*_{\mu_1\cdots\mu_n}&=\msR_{\mu_1\cdots\mu_n},~~~
\label{BRST phi* AdS}\\
\delta_B \zeta^*_{\mu_2\cdots\mu_n}&=
-n\hat\msD\cdot\psi^*_{\mu_2\cdots\mu_n}
+
\frac{n}{D+2n-4}
(\gamma\hat\msD\cdot\psisl^*
+\gb \hat\msD\cdot\psis{}')
,
\label{BRST zeta* AdS}
\end{align}
when ${D+2n-4}\neq 0$\footnote{When $D+2n-4=0$, namely $(D,n)=(4,0)$,
the gauge transformation \bref{eqn:gauge trans fermion AdS} becomes trivial.
}.
We have added the second term on the right-hand side of \bref{BRST zeta* AdS}
for the nilpotency $\delta_B^2 \zeta^*=0$.
We find that the key relation for the nilpotency
\begin{align}
-\hat\msD\cdot \msR
+
\frac{1}{D+2n-4}
(\gamma\hat\msD\cdot\msRsl
+\gb \hat\msD\cdot\msR')
=0
\label{eqn: fer nilpotency AdS}
\end{align}
follows from the triple $\gamma$-traceless condition $\psisl'=0$.
Since $\delta_B \zeta^*=(\zeta^*,S^0)$,
the second term on the right-hand side of \bref{BRST zeta* AdS}
requires an additional term
\begin{align}
\frac{n}{D+2n-4}\int \d^Dx\sqrt{-\gb}\left(
\bar\zeta(\gamma\hat\msD\cdot\psisl^*+\gb\hat\msD\cdot\psis')
-\overline{(\gamma\hat\msD\cdot\psisl^*+\gb\hat\msD\cdot\psis{}')}\zeta
\right)
\,,
\end{align}
in the action $S^0$.
However this term disappears due to $\zetasl=0$ \footnote{$\xi'=0$ follows from $\zetasl=0$},
and leaves the action \bref{eqn:S^0 psi AdS} unchanged.

\medskip

We have introduced free fermionic  higher-spin gauge theory
on AdS spaces
in the BRST-antifield formalism.
In appendix  \ref{sec:bos AdS},
the free bosonic higher-spin gauge theory on AdS spaces
is introduced.
The action of $\Delta$ and $\Gamma$ on fields and antifields is summarized in Table \ref{table:Delta and Gamma AdS}.
$\msG$ and $\msR$ are defined in \bref{eqn:G AdS} and \bref{eqn:R AdS}, respectively.
\begin{table}[htb]
\begin{align*}
\begin{array}{ccc}\hline 
Z & \Delta(Z) & \Gamma(Z) \\\hline 
\phi_{\mu_1\cdots\mu_{n}} & 0 & \nabla_{(\mu_1}c_{\mu_2\cdots\mu_{n})} \\
c_{\mu_2\cdots\mu_{n}} & 0 & 0  \\
\phi^*_{\mu_1\cdots\mu_{n}} & -\msG_{\mu_1\cdots\mu_{n}}(\phi) & 0  \\
\disp 
c^*_{\mu_2\cdots\mu_{n}} 
&
 -{n}\nabla\cdot\phi^*_{\mu_2\cdots\mu_{n}} 
+\frac{{n}}{D+2{n}-6}
\gb_{(\mu_2\mu_3}\nabla\cdot \phi^*{}'_{\mu_4\cdots\mu_n)}
& 0  \\\hline 
\psi_{\mu_1\cdots\mu_n} & 0 & \hat\msD_{(\mu_1}\zeta_{\mu_2\cdots\mu_n)} \\
\zeta_{\mu_2\cdots\mu_{n}} & 0 & 0  \\
\psi^*_{\mu_1\cdots\mu_n} & \msR_{\mu_1\cdots\mu_n} & 0  \\
\zeta^*_{\mu_2\cdots\mu_{n}} &
 -n\hat\msD\cdot\psi^*_{\mu_2\cdots\mu_n} 
+\frac{n}{D+2n-4}(\gamma_{(\mu_2}\hat\msD\cdot \psisl^*_{\mu_3\cdots\mu_n)}
+\gb_{(\mu_2\mu_3}\hat\msD\cdot \psi^*{}'_{\mu_4\cdots\mu_n)})
& 0  \\\hline 
\end{array}
\end{align*}
 \caption{Action of $\Delta$ and $\Gamma$}
 \label{table:Delta and Gamma AdS}
\end{table}

Vertices can be constructed by using the BRST deformation scheme,
as was done in sections \ref{sec:BRST deformation}, \ref{sec:three fields}
and \ref{sec:two fields}.
We will not repeat the procedure here but present the results below.
The deformed action of tensor-spinors $\psi_I$ ($I=1,2$) and a tensor $\phi$
is found to be\footnote{The determinant of the metric, $\gb$, may not be confused with the coupling $g$.}
\begin{align}
S=&S_0+S_1+S_2,\\
S_0=&\intd\sqrt{-\gb}\,\tr\bigg[
\frac{1}{2}\phi^T\msG(\phi)
-\bar\psi_I
\sum_{k=0}^\infty[(igs)^k]^{IJ}\varPsi^k[\msR(\psi_J)]
\bigg],
\label{eqn:S0 three fer AdS}\\
S_1=&\intd\sqrt{-\gb}\,
\tr\bigg[
\phi^*{}^T\nabla c
+\bar\psi^*_I \hat\msD\zeta_I
-\overline{\hat\msD\zeta}_I \psi^*_I
\nn\\&\hspace{13mm}
+\bar\psi^*_I\tilde\nabla c 
\sum_{k=0}^\infty[(igs)^{k+1}]^{IJ}\varPsi^k
[\msR(\psi_J)]
-
\sum_{k=0}^\infty \overline{[(igs)^{k+1}]^{IJ}\varPsi^k
[\msR(\psi_J)]
}\tilde\nabla c^T\psi^*_I
\bigg],\\
S_2=&\intd\sqrt{-\gb}\,\tr
\bar\psi^*_I \tilde\nabla c 
\sum_{k=0}^\infty[(igs)^{k+2}]^{IJ}\varPsi^k
[\msRt(\tilde\nabla c^T \psi^*_J)],
\end{align}
where $\varPsi$ is defined as
$\varPsi^0[A]=A$,
$\varPsi^1[A]=\msRt(\phi A)$,
$\varPsi^2[A]=\msRt(\phi \msRt(\phi A))$,
and so on.
To derive this result,  we frequently used the relation
\begin{align}
\intd \sqrt{-\gb}\,\bar A \varPsi^k[\msRt(B)]=
(-1)^{k+1}
\intd \sqrt{-\gb}\,\overline{\varPsi^k[\msRt(A)]}B.
\end{align}
The obtained action reduces to \bref{eqn:S 3 agh AdS}-\bref{eqn:S2 3 agh AdS}
in the flat limit as expected.

Furthermore we obtain the deformed action of a tensor-spinor $\psi$ and a tensor $\phi$
\begin{align}
S=&S_0+S_1+S_2,\\
S_0=&\intd\sqrt{-\gb}\,
\tr\bigg[
\frac{1}{2}\phi^T\msG(\phi)
-\bar\psi
\sum_{k=0}^\infty(ig)^k\varPsi^k
[\msR(\psi)]
\bigg],
\label{eqn:S0 two fer AdS}\\
S_1=&\intd\sqrt{-\gb}\,\tr\bigg[
\phi^*{}^T\nabla c
+\bar\psi^* \hat\msD\zeta
-\overline{\hat\msD\zeta} \psi^*
\nn\\&\hspace{13mm}
+\bar\psi^*\tilde\nabla c 
\sum_{k=0}^\infty(ig)^{k+1}\varPsi^k
[\msR(\psi)]
-
\sum_{k=0}^\infty \overline{(ig)^{k+1}\varPsi^k
[\msR(\psi)]
}\,\tilde\nabla c^T\psi^*
\bigg],\\
S_2=&\intd\sqrt{-\gb}\,\tr
\bar\psi^*\tilde\nabla c 
\sum_{k=0}^\infty(ig)^{k+2}\varPsi^k
[\msRt(\tilde\nabla c^T \psi^*)]
.
\end{align}
This action reduces to \bref{eqn:S 2 AdS}-\bref{eqn:S2 2 AdS}
in the flat limit as expected.
We note that
these actions contain terms of all orders in $g$,
and are invariant exactly under the BRST-transformation $\delta_B^g S=(S,S)=0$.
We have obtained BRST-invariant deformed actions of  gauge fields
on AdS spaces
by applying the BRST deformation scheme.

We can show that the actions \bref{eqn:S0 three fer AdS} and \bref{eqn:S0 two fer AdS} reduce to free on-shell
 in a similar way as in the flat case.
Furthermore,
we find that the action $S_0$ in \bref{eqn:S0 three fer AdS}
is invariant under the gauge transformations
with  rank-$(n_I-1)$ tensor-spinor parameters $\epsilon_I$
\begin{align}
\delta \phi=0,~~~
\delta \psi_I=\hat \msD\epsilon_I,
\end{align}
and under the gauge transformation with  a rank-$(n-1)$ tensor parameter $\xi$ 
\begin{align}
\delta \phi=\nabla \xi,~~~
\delta \psi_I=\tilde\nabla \xi
\sum_{k=0}^\infty[(igs)^{k+1}]^{IJ}\varPsi^k
[\msR(\psi_J)].
\label{eqn:gauge trans 3 fermionic AdS}
\end{align}
It is also shown that
the action \bref{eqn:S0 two fer AdS}
is invariant under the gauge transformation
with  a rank-$(n_1-1)$ tensor-spinor  parameter $\epsilon$
\begin{align}
\delta \phi=0,~~~
\delta \psi=\hat \msD\epsilon,
\end{align}
and under the gauge transformation with  a rank-$(n-1)$ tensor parameter $\xi$ 
\begin{align}
\delta \phi=\nabla \xi,~~~
\delta \psi=\tilde\nabla \xi
\sum_{k=0}^\infty(ig)^{k+1}\varPsi^k
[\msR(\psi)].
\label{eqn:gauge trans 2 fermionic AdS}
\end{align}
We note that the gauge transformations \bref{eqn:gauge trans 3 fermionic AdS}
and \bref{eqn:gauge trans 2 fermionic AdS} are deformed by vertices,
but the gauge algebra remains abelian on-shell.

\section{Summary and Discussion}\label{summary}

Introducing totally-symmetric rank-$n$ tensor-spinors,
which are Dirac spinors in $D$-dimensional spacetime,
as well as totally-symmetric rank-$n$ tensors,
we constructed higher-spin gauge models including fermions
by applying the BRST deformation scheme.
The deformed action $S$ contains terms of all orders in the deformation parameter $g$
and satisfy the master equation exactly.
Introducing objects on AdS spaces
and writing down the BRST transformation law on AdS spaces in the BRST-antifield formalism,
we applied the BRST deformation scheme
to derive higher-spin gauge models on AdS-spaces.

Extending the case \cite{SS2011}  that each vertex forms an open chain of fields,
we constructed the models in which each vertex forms closed chain of fields.
The models presented here contain the previous models as special cases.
These results are summarized in appendices.

\medskip

The actions obtained in this paper contain  infinite series,
say $\sum_{k=1}^\infty [(igs)^k]^{IJ}\Psi^k[\CR(\psi_J)]$
in \bref{eqn:S infinite}.
This term is expressed  in a closed form $(\frac{1}{1-igs\Psi})^{IJ}\CR(\psi_J)$
assuming $|igs\Psi|<1$.
It is not clear what this condition means,
as $\Psi$ contains a tensor and differentials.
However,
let us assume this condition satisfied anyway.
In this case, the equation of motion \bref{eqn:eom fermion}
is expressed as
$(\frac{1}{1-igs\Psi})^{IJ}\CR(\psi_J)=0$.
By 
applying $({1-igs\Psi})^{IJ}$,
this reduces formally to $\CR(\psi_J)=0$,
which is the free equation of motion formally.
For bosonic gauge models,
infinite series, say $\sum_{k=0}^\infty [(gs)^k]^{IJ}\Phi^k[G(\phi_J)]$ in \bref{eqn:infinite bosonic},
is expressed as a closed form,
$(\frac{1}{1-gs\Phi})^{IJ}G(\phi_J)$,
assuming $|gs\Phi|<1$.
In this case,
the equation of motion \bref{eqn:eom boson}
is expressed as $(\frac{1}{1-gs\Phi})^{IJ}G(\phi_J)=0$,
and reduces to $G(\phi_J)=0$.
As a result,
the higher-spin gauge models may reduce formally to free theory on-shell
when infinite series contained in these models are convergent.

Furthermore,
we can show that
the higher-spin gauge models up to the cubic vertex reduce to free theory even off-shell.
When we construct the deformed action,
we choose $S^1$ to satisfy the master equations at the first order.
As $\alpha_2^1$ can be absorbed into $S^0$ by field redefinitions,
we set $\alpha_2^1=0$.
But, this may not be enough to avoid a BRST-exact $S^1$.
In fact,
$S^1$ is shown to be BRST-exact:
\begin{align}
S^1=(R,S^0),
\end{align}
where,
for example,
\begin{align}
R=-\frac{1}{2}\intd \tr\Big[
\phi_1^*{}^T\tilde\partial c\phi_1^*
+\phi_1^*{}^T\phi G(\phi_1)
\Big],
\end{align}
in the case with two bosons on flat space.
So $S^1$ can be absorbed into $S^0$ by field redefinitions. 
However, field redefinitions to reduce the total action $S$  to the free action $S^0$
are not yet known.
Whether the higher spin gauge models are free off-shell or not is a non-trivial question 
and is left as a future problem.
We will report on this issue in \cite{FS2024}.

Our models contain the deformation parameter $g$.
As the dimension of a tensor $\phi$ is $\frac{D-2}{2}$ and that of a tensor-spinor $\psi$ is $\frac{D-1}{2}$,
the mass dimension of $g$
is found to be $-\frac{D+2}{2}$ for bosonic models given in appendices
while
$-\frac{D}{2}$ for models including fermions.
Since the dimensions of $g\Phi$ and $g\Psi$ are zero,
vertices have the same dimension as expected.

Our models are BRST-invariant up to surface terms.
We have dropped surface terms assuming that there is no boundary.
Boundary terms in AdS spaces are examined in \cite{JM1112}.
It is interesting to consider a boundary action such that the total action including the boundary action
is BRST-invariant.

There are many interesting issues, such as including massive tensors and massive tensor-spinors,
including tensors and tensor-spinors
with mixed-symmetry,
and constructing the unconstrained local version \cite{FS0207,FS0212,FS0507,FMS0701} of our models.
These are left for future investigation.

\medskip

\section*{Acknowledgments}

The authors would like to thank Takanori Fujiwara, Yoshifumi Hyakutake,
 and Sota Hanazawa
for useful comments. 
This work was supported by JSPS KAKENHI Grant Number
JP21K03566
and
JST,
the establishment of university fellowships towards the creation of science technology innovation, 
Grant Number JPMJFS2105.

\appendix
\section*{Appendix}

\section{Higher-spin gauge model of two bosons}\label{sec:two bosonic gauge fields}

We present a deformed higher-spin gauge model of two bosonic tensors,
which is a generalization of the model given in \cite{SS2011}.
We introduce two gauge fields,
rank-$n_1$ tensor  $\phi_1$
and a rank-$n$ tensor $\phi$.
We also introduce ghosts $c_1$ and $c$, 
and antifields $\{\phi^*_1,\phi^*,c^*_1,c^*\}$.
The free action is 
\begin{align}
S^0=
\intd
\bigg[&
\frac{1}{2}\phi_1^{\mu_1\cdots\mu_{n_1}}G_{\mu_1\cdots\mu_{n_1}}(\phi_1)
+\phi^*_1{}^{\mu_1\cdots \mu_{n_1}}\pa_{(\mu_1}c_1{}_{\mu_2\cdots \mu_{n_1})}
\nn\\&
+\frac{1}{2}\phi^{\mu_1\cdots\mu_{n}}G_{\mu_1\cdots\mu_{n}}(\phi)
+\phi^*{}^{\mu_1\cdots \mu_{n}}\pa_{(\mu_1}c_{\mu_2\cdots \mu_{n})}
\bigg].
\label{eqn:action 3phi}
\end{align}
In the manner similar to one explained in appendix \ref{sec:three bosonic gauge firleds},
we can show that the higher order term
is given as
$ S^n=\alpha_2^n+\alpha_1^n+\alpha_0^n$
with
\begin{align}
\alpha_2^n=&\intd \frac{1}{2}\tr (\pat c \phi_1^*)^T \Phi^{n-2}[\Gt(\pat c \phi^*_1)],
\label{eqn:Sn alpha2 bos2}\\
\alpha_1^n=&
\intd  \tr (\pat c \phi_1^*)^T \Phi^{n-1}[\Gt(\phi_1)],
\label{eqn:Sn alpha1 bos2}\\
\alpha_0^n=&
\intd  \frac{1}{2}\tr \phi_1^T \Phi^{n}[\Gt(\phi_1)].
\label{eqn:Sn alpha0 bos2}
\end{align}
We find that the action turns to the form
expanded in $agh$
as
\begin{align}
S=&S_0+S_1+S_2,\\
S_0=&\intd\tr\Big[
\frac{1}{2} \phi_1^T 
\sum_{k=0}^\infty g^k\Phi^k
[G(\phi_1)]
+\frac{1}{2}\phi^TG(\phi)
\Big],\\
S_1=&\intd \tr \Big[
\phi_1^*{}^T\pa c_1
+\phi^*{}^T\pa c
+\phi_1^T
\sum_{k=0}^\infty g^{k+1}\Phi^k
[\Gt(\pat c \phi_1^*)]
\Big],\\
S_2=&\intd \tr
\frac{1}{2}(\pat c\phi_1^*)^T
\sum_{k=0}^\infty g^{k+2}\Phi^k
[\Gt(\pat c \phi_1^*)].
\end{align}
We can show that $S_0$ leads to field equations of
$G(\phi)=0$ and $G(\phi_1)=0$
assuming $|g\Phi|<1$.
The action $S_0$
is invariant under the gauge transformation with a rank-$(n_1-1)$ tensor parameter $\xi_1$
\begin{align}
\delta \phi=0,
~~~
\delta \phi_1=\pa \xi_1,
\end{align}
and the gauge transformation with  a rank-$(n-1)$ tensor  parameter $\xi$
\begin{align}
\delta \phi=\pa\xi,
~~~
\delta \phi_1=-
\pat \xi^T
\sum_{k=0}^\infty g^{k+1}\Phi^k
[G(\phi_1)].
\end{align}

It is straightforward to derive
the corresponding action on AdS spaces 
\begin{align}
S=&S_0+S_1+S_2,\\
S_0=&\intd\sqrt{-\gb}\,\tr\Big[
\frac{1}{2} \phi_1^T 
\sum_{k=0}^\infty g^k\varPhi^k
[\msG(\phi_1)]
+\frac{1}{2}\phi^T \msG(\phi)
\Big],
\label{eqn:action AdS 2bos S0}\\
S_1=&\intd\sqrt{-\gb}\, \tr \Big[
\phi_1^*{}^T\nabla c_1
+\phi^*{}^T\nabla c
+\phi_1^T 
\sum_{k=0}^\infty g^{k+1}\varPhi^k
[\msGt(\tilde\nabla c \phi_1^*)]
\Big],\\
S_2=&\intd\sqrt{-\gb}\, \tr
\frac{1}{2}(\tilde\nabla c\phi_1^*)^T
\sum_{k=0}^\infty g^{k+2}\varPhi^k
[\msGt(\tilde\nabla c \phi_1^*)].
\end{align}
Again we find that varying $S_0$ leads to
$\msG(\phi)=0$ and $\msG(\phi_1)=0$
assuming $|g\varPhi|<1$.
The action $S_0$ in  \bref{eqn:action AdS 2bos S0}
is invavriant under 
the gauge transformation with a rank-$(n_1-1)$ tensor parameter $\xi_1$
\begin{align}
\delta \phi=0,~~~
\delta \phi_1=\nabla \xi_1,
\end{align}
and
the gauge transformation with  a rank-$(n-1)$ tensor parameter $\xi$ 
turns to
\begin{align}
\delta \phi=\nabla\xi,~~~
\delta \phi_1=-
\tilde\nabla \xi^T
\sum_{k=0}^\infty g^{k+1}\varPhi^k
[\msG(\phi_1)].
\end{align}
The  gauge algebra remains abelian on-shell.


\begin{thebibliography}{99}




\bibitem{Gross}
D.~J.~Gross,
``High-Energy Symmetries of String Theory,''
Phys. Rev. Lett. \textbf{60} (1988), 1229


\bibitem{Sagnotti tensionless limit}
A.~Sagnotti and M.~Taronna,
``String Lessons for Higher-Spin Interactions,''
Nucl. Phys. B \textbf{842} (2011), 299-361
[arXiv:1006.5242 [hep-th]].



\bibitem{AdS/CFT}
J.~M.~Maldacena,
``The Large N limit of superconformal field theories and supergravity,''
Int. J. Theor. Phys. \textbf{38} (1999), 1113-1133,
 Adv. Theor. Math. Phys. 2 (1998) 231-252
[arXiv:hep-th/9711200 [hep-th]].


S.~S.~Gubser, I.~R.~Klebanov and A.~M.~Polyakov,
``Gauge theory correlators from noncritical string theory,''
Phys. Lett. B \textbf{428} (1998), 105-114
[arXiv:hep-th/9802109 [hep-th]].


E.~Witten,
``Anti-de Sitter space and holography,''
Adv. Theor. Math. Phys. \textbf{2} (1998), 253-291
[arXiv:hep-th/9802150 [hep-th]].






\bibitem{3dVasiliev}
S.~F.~Prokushkin and M.~A.~Vasiliev,
``Higher spin gauge interactions for massive matter fields in 3-D AdS space-time,''
Nucl. Phys. B \textbf{545} (1999), 385
[arXiv:hep-th/9806236 [hep-th]];
``3-d higher spin gauge theories with matter,''
[arXiv:hep-th/9812242 [hep-th]].

\bibitem{3dV/W}
M.~R.~Gaberdiel and R.~Gopakumar,
``An AdS$_{3}$ Dual for Minimal Model CFTs,''
Phys. Rev. D \textbf{83} (2011), 066007
[arXiv:1011.2986 [hep-th]].



\bibitem{4dVasiliev}
M.~A.~Vasiliev,
``Nonlinear equations for symmetric massless higher spin fields in (A)dS(d),''
Phys. Lett. B \textbf{567} (2003), 139-151
[arXiv:hep-th/0304049 [hep-th]].



\bibitem{4dV/O}
I.~R.~Klebanov and A.~M.~Polyakov,
``AdS dual of the critical O(N) vector model,''
Phys. Lett. B \textbf{550} (2002), 213-219
[arXiv:hep-th/0210114 [hep-th]].






\bibitem{no-go}
S.~Weinberg,
``Photons and Gravitons in  $S$-Matrix Theory: Derivation of Charge Conservation and Equality of Gravitational and Inertial Mass,''
Phys. Rev. \textbf{135} (1964), B1049-B1056.

\bibitem{KU81}
T.~Kugo and S.~Uehara,
``Massless Particle With Spin $J \geq 1$ Implies the S-Matrix Symmetry''
Prog. Theor. Phys. \textbf{66} (1981), 1044.







\bibitem{Met91}
R.~R.~Metsaev,
``Poincare invariant dynamics of massless higher spins: Fourth order analysis on mass shell,''
Mod. Phys. Lett. A \textbf{6} (1991), 359-367
;
``S matrix approach to massless higher spins theory. 2: The Case of internal symmetry,''
Mod. Phys. Lett. A \textbf{6} (1991), 2411-2421


\bibitem{PS1609}
D.~Ponomarev and E.~D.~Skvortsov,
``Light-Front Higher-Spin Theories in Flat Space,''
J. Phys. A \textbf{50} (2017) no.9, 095401
[arXiv:1609.04655 [hep-th]].

\bibitem{STT2002}
E.~Skvortsov, T.~Tran and M.~Tsulaia,
``More on Quantum Chiral Higher Spin Gravity,''
Phys. Rev. D \textbf{101} (2020) no.10, 106001
[arXiv:2002.08487 [hep-th]].


\bibitem{ST2004}
E.~Skvortsov and T.~Tran,
``One-loop Finiteness of Chiral Higher Spin Gravity,''
JHEP \textbf{07} (2020), 021
[arXiv:2004.10797 [hep-th]].




\bibitem{P1710}
D.~Ponomarev,
``Chiral Higher Spin Theories and Self-Duality,''
JHEP \textbf{12} (2017), 141
[arXiv:1710.00270 [hep-th]].


\bibitem{KST2105}
K.~Krasnov, E.~Skvortsov and T.~Tran,
``Actions for Self-dual Higher Spin Gravities,''
[arXiv:2105.12782 [hep-th]].





\bibitem{Metsaev0512}
R.~R.~Metsaev,
``Cubic interaction vertices of massive and massless higher spin fields,''
Nucl. Phys. B \textbf{759} (2006), 147-201
[arXiv:hep-th/0512342 [hep-th]].

\bibitem{Metsaev0712}
R.~R.~Metsaev,
``Cubic interaction vertices for fermionic and bosonic arbitrary spin fields,''
Nucl. Phys. B \textbf{859} (2012), 13-69
[arXiv:0712.3526 [hep-th]].


\bibitem{Bonora2020}
L.~Bonora and S.~Giaccari,
``Supersymmetric HS Yang-Mills-like models,"
Universe \textbf{6} (2020) 12, 245
[arXive:2011.00734[hep-th]];
``HS in flat spacetime. YM-like models,"
[arXive:1812.05030[hep-th]]. 




\bibitem{BBvD85}
F.~A.~Berends, G.~J.~H.~Burgers and H.~van Dam,
``On the Theoretical Problems in Constructing Interactions Involving Higher Spin Massless Particles,''
Nucl. Phys. B \textbf{260} (1985), 295-322.




\bibitem{MMR Noether}
R.~Manvelyan, K.~Mkrtchyan and W.~R\"uhl,
``Off-shell construction of some trilinear higher spin gauge field interactions,''
Nucl. Phys. B \textbf{826} (2010), 1-17
[arXiv:0903.0243 [hep-th]];
``A Generating function for the cubic interactions of higher spin fields,''
Phys. Lett. B \textbf{696} (2011), 410-415
[arXiv:1009.1054 [hep-th]].

K.~Mkrtchyan,
``On generating functions of Higher Spin cubic interactions,''
Phys. Atom. Nucl. \textbf{75} (2012), 1264-1267
[arXiv:1101.5643 [hep-th]].


\bibitem{Sleight1704}
C.~Sleight and M.~Taronna,
``Higher-Spin Gauge Theories and Bulk Locality,''
Phys. Rev. Lett. \textbf{121} (2018) no.17, 171604
[arXiv:1704.07859 [hep-th]].


\bibitem{KMP2104}
M.~Karapetyan, R.~Manvelyan and G.~Poghosyan,
``On special quartic interaction of higher spin gauge fields with scalars and gauge symmetry commutator in the linear approximation,''
Nucl. Phys. B \textbf{971} (2021), 115512
[arXiv:2104.09139 [hep-th]].





\bibitem{PT9803}
A.~Pashnev and M.~Tsulaia,
``Description of the higher massless irreducible integer spins in the BRST approach,''
Mod. Phys. Lett. A \textbf{13} (1998), 1853-1864
[arXiv:hep-th/9803207 [hep-th]].


\bibitem{BPT0101}
C.~Burdik, A.~Pashnev and M.~Tsulaia,
``On the Mixed symmetry irreducible representations of the Poincare group in the BRST approach,''
Mod. Phys. Lett. A \textbf{16} (2001), 731-746
[arXiv:hep-th/0101201 [hep-th]];
``The Lagrangian description of representations of the Poincare group,''
Nucl. Phys. B Proc. Suppl. \textbf{102} (2001), 285-292
[arXiv:hep-th/0103143 [hep-th]].

\bibitem{BKP0410}
I.~L.~Buchbinder, V.~A.~Krykhtin and A.~Pashnev,
``BRST approach to Lagrangian construction for fermionic massless higher spin fields,''
Nucl. Phys. B \textbf{711} (2005), 367-391
[arXiv:hep-th/0410215 [hep-th]].



\bibitem{BK0505}
I.~L.~Buchbinder and V.~A.~Krykhtin,
``Gauge invariant Lagrangian construction for massive bosonic higher spin fields in D dimensions,''
Nucl. Phys. B \textbf{727} (2005), 537-563
[arXiv:hep-th/0505092 [hep-th]];
``BRST approach to higher spin field theories,''
[arXiv:hep-th/0511276 [hep-th]].


\bibitem{BKRT0603}
I.~L.~Buchbinder, V.~A.~Krykhtin, L.~L.~Ryskina and H.~Takata,
``Gauge invariant Lagrangian construction for massive higher spin fermionic fields,''
Phys. Lett. B \textbf{641} (2006), 386-392
[arXiv:hep-th/0603212 [hep-th]].


\bibitem{BR2010}
\v{C}.~Burd\'\i{}k and A.~A.~Reshetnyak,
``BRST-BV Quantum Actions for Constrained Totally-Symmetric Integer HS Fields,''
Nucl. Phys. B \textbf{965} (2021), 115357
[arXiv:2010.15741 [hep-th]].


\bibitem{BKTW2103}
I.~L.~Buchbinder, V.~A.~Krykhtin, M.~Tsulaia and D.~Weissman,
``Cubic Vertices for N=1 Supersymmetric Massless Higher Spin Fields in Various Dimensions,''
Nucl. Phys. B \textbf{967} (2021), 115427
[arXiv:2103.08231 [hep-th]].


\bibitem{BR2105}
I.~L.~Buchbinder and A.~A.~Reshetnyak,
``General cubic interacting vertex for massless integer higher spin fields,''
Phys. Lett. B \textbf{820} (2021), 136470
[arXiv:2105.12030 [hep-th]].





\bibitem{BFPT0609}
I.~L.~Buchbinder, A.~Fotopoulos, A.~C.~Petkou and M.~Tsulaia,
``Constructing the cubic interaction vertex of higher spin gauge fields,''
Phys. Rev. D \textbf{74} (2006), 105018
[arXiv:hep-th/0609082 [hep-th]].



\bibitem{FIPT0708}
A.~Fotopoulos, N.~Irges, A.~C.~Petkou and M.~Tsulaia,
``Higher-Spin Gauge Fields Interacting with Scalars: The Lagrangian Cubic Vertex,''
JHEP \textbf{10} (2007), 021
[arXiv:0708.1399 [hep-th]].




\bibitem{BCS0409}
N.~Bouatta, G.~Compere and A.~Sagnotti,
``An Introduction to free higher-spin fields,''
[arXiv:hep-th/0409068 [hep-th]].

\bibitem{FT0805}
A.~Fotopoulos and M.~Tsulaia,
``Gauge Invariant Lagrangians for Free and Interacting Higher Spin Fields. A Review of the BRST formulation,''
Int. J. Mod. Phys. A \textbf{24} (2009), 1-60
[arXiv:0805.1346 [hep-th]].












\bibitem{BH9304}
G.~Barnich and M.~Henneaux,
``Consistent couplings between fields with a gauge freedom and deformations of the master equation,''
Phys. Lett. B \textbf{311} (1993), 123-129
[arXiv:hep-th/9304057 [hep-th]].


M.~Henneaux,
``Consistent interactions between gauge fields: The Cohomological approach,''
Contemp. Math. \textbf{219} (1998), 93-110
[arXiv:hep-th/9712226 [hep-th]].






\bibitem{BDGH0007}
  N.~Boulanger, T.~Damour, L.~Gualtieri and M.~Henneaux,
  ``Inconsistency of interacting, multigraviton theories,''
  Nucl.\ Phys.\ B {\bf 597} (2001) 127
  [hep-th/0007220].




\bibitem{BRST cohomology}
N.~Boulanger and S.~Leclercq,
``Consistent couplings between spin-2 and spin-3 massless fields,''
JHEP \textbf{11} (2006), 034
[arXiv:hep-th/0609221 [hep-th]].

N.~Boulanger, S.~Leclercq and P.~Sundell,
``On The Uniqueness of Minimal Coupling in Higher-Spin Gauge Theory,''
JHEP \textbf{08} (2008), 056
[arXiv:0805.2764 [hep-th]].






\bibitem{HLGR1206}
M.~Henneaux, G.~Lucena G\'omez and R.~Rahman,
``Higher-Spin Fermionic Gauge Fields and Their Electromagnetic Coupling,''
JHEP \textbf{08} (2012), 0933
[arXiv:1206.1048 [hep-th]].


\bibitem{HLGR1310}
M.~Henneaux, G.~Lucena G\'omez and R.~Rahman,
``Gravitational Interactions of Higher-Spin Fermions,''
JHEP \textbf{01} (2014), 087
[arXiv:1310.5152 [hep-th]].

\bibitem{Rahman1905}
R.~Rahman,
``The Uniqueness of Hypergravity,''
JHEP \textbf{11} (2019), 115
[arXiv:1905.04109 [hep-th]].





\bibitem{BL2104}
I.~L.~Buchbinder and P.~M.~Lavrov,
``On a gauge-invariant deformation of a classical gauge-invariant theory,''
JHEP \textbf{06} (2021), 097
[arXiv:2104.11930 [hep-th]]
;
``On classical and quantum deformations of gauge theories,''
[arXiv:2108.09968 [hep-th]].


M.~Taronna,
``Higher Spins and String Interactions,''
[arXiv:1005.3061 [hep-th]].









\bibitem{ST0311}
A.~Sagnotti and M.~Tsulaia,
``On higher spins and the tensionless limit of string theory,''
Nucl. Phys. B \textbf{682} (2004), 83-116
[arXiv:hep-th/0311257 [hep-th]].

D.~Francia and A.~Sagnotti,
``On the geometry of higher spin gauge fields,''
Class. Quant. Grav. \textbf{20} (2003), S473-S486
[arXiv:hep-th/0212185 [hep-th]].






\bibitem{FT0705}
A.~Fotopoulos and M.~Tsulaia,
``Interacting higher spins and the high energy limit of the bosonic string,''
Phys. Rev. D \textbf{76} (2007), 025014
[arXiv:0705.2939 [hep-th]].

\bibitem{FT1009}
A.~Fotopoulos and M.~Tsulaia,
``On the Tensionless Limit of String theory, Off - Shell Higher Spin Interaction Vertices and BCFW Recursion Relations,''
JHEP \textbf{11} (2010), 086
[arXiv:1009.0727 [hep-th]].




\bibitem{Polyakov0910}
D.~Polyakov,
``Interactions of Massless Higher Spin Fields From String Theory,''
Phys. Rev. D \textbf{82} (2010), 066005
[arXiv:0910.5338 [hep-th]]
;
``Gravitational Couplings of Higher Spins from String Theory,''
Int. J. Mod. Phys. A \textbf{25} (2010), 4623-4640
arXiv:1005.5512 [hep-th]].


\bibitem{Sagnotti1112}
A.~Sagnotti,
``Notes on Strings and Higher Spins,''
J. Phys. A \textbf{46} (2013), 214006
[arXiv:1112.4285 [hep-th]].


\bibitem{SS2011}
M.~Sakaguchi and H.~Suzuki,
``On interacting higher-spin bosonic gauge fields in the BRST-antifield formalism,''
Prog.~Theor.~Exp.~Phys. \textbf{2021} (2021) no.4, 043B01
[arXiv:2011.02689 [hep-th]].




\bibitem{Fronsdal78}
C.~Fronsdal,
``Massless Fields with Integer Spin,''
Phys. Rev. D \textbf{18} (1978), 3624.




\bibitem{FangFronsdal78}
J.~Fang and C.~Fronsdal,
``Massless Fields with Half Integral Spin,''
Phys. Rev. D \textbf{18} (1978), 3630
doi:10.1103/PhysRevD.18.3630




\bibitem{FS2024}
R.~Fujii and M.~Sakaguchi,
``A Comment on the Higher-Spin Gauge Models"
Mod.~Phys.~Lett. A  (2025) 2550007,
doi:10.1142/S0217732325500075
[arXiv:2411.13984 [hep-th]].









\bibitem{JM1112}
E.~Joung and J.~Mourad,
``Boundary action of free AdS higher-spin gauge fields and the holographic correspondence,''
JHEP \textbf{06} (2012), 161
[arXiv:1112.5620 [hep-th]].




\bibitem{FS0207}
D.~Francia and A.~Sagnotti,
``Free geometric equations for higher spins,''
Phys. Lett. B \textbf{543} (2002), 303-310
[arXiv:hep-th/0207002 [hep-th]].

\bibitem{FS0212}
D.~Francia and A.~Sagnotti,
``On the geometry of higher spin gauge fields,''
Class. Quant. Grav. \textbf{20} (2003), S473-S486
[arXiv:hep-th/0212185 [hep-th]].



\bibitem{FS0507}
D.~Francia and A.~Sagnotti,
``Minimal local Lagrangians for higher-spin geometry,''
Phys. Lett. B \textbf{624} (2005), 93-104
[arXiv:hep-th/0507144 [hep-th]].

\bibitem{FMS0701}
D.~Francia, J.~Mourad and A.~Sagnotti,
``Current Exchanges and Unconstrained Higher Spins,''
Nucl. Phys. B \textbf{773} (2007), 203-237
[arXiv:hep-th/0701163 [hep-th]].








\end{thebibliography}
\end{document}